\documentclass[aps,prb,reprint,showpacs,superscriptaddress,longbibliography]{revtex4-1}
\usepackage{mathtools,amssymb,graphicx,units,multirow}
\usepackage[usenames,dvipsnames]{color}
\usepackage[plainpages=false,pdfpagelabels,colorlinks=true,linkcolor=red,urlcolor=blue,citecolor=blue,pdftitle={Title},pdfauthor={},pdfdisplaydoctitle=true,pdfduplex=DuplexFlipLongEdge]{hyperref}

\usepackage{booktabs}

\newcommand{\ra}[1]{\renewcommand{\arraystretch}{#1}}
\graphicspath{{./figures/}}
\begin{document}

\title{Edge and bulk localization of Floquet topological superconductors}

\author{Tilen \v Cade\v z}
\email{tilen.cadez@csrc.ac.cn}
\affiliation{Beijing Computational Science Research Center, Beijing, 100193, China}
\author{Rubem Mondaini}
\email{rmondaini@csrc.ac.cn}
\affiliation{Beijing Computational Science Research Center, Beijing, 100193, China}
\author{Pedro D. Sacramento}
\email{pdss@cefema.tecnico.ulisboa.pt}
\affiliation{CeFEMA, Instituto Superior T\'{e}cnico, Universidade de Lisboa, Av. Rovisco Pais, 1049-001 Lisboa, Portugal}
\affiliation{Beijing Computational Science Research Center, Beijing, 100193, China}

\begin{abstract}
We study the bulk and edge properties of a driven Kitaev chain, where the driving is performed as instantaneous quenches of the on-site energies. We identify three periodic driving regimes: low period, which is equivalent to a static model, with renormalized parameters obtained from the Baker-Campbell-Hausdorff (BCH) expansion; intermediate period, where the first order BCH expansion breaks down; and high period when the quasienergy gap at $\omega/2$ closes. We investigate the dynamical localization properties for the case of quasiperiodic potential driving as a function of its amplitude and the pairing strength, obtaining regimes with extended, critical and localized bulk states, if the driving is performed at high frequencies. In these, we characterize wave-packet propagation, obtaining ballistic, subdiffusive and absence of spreading, respectively. In the intermediate period regime, we find an additional region in the phase diagram with a mobility edge between critical and localized states. Further, we investigate the stability of these phases under time-aperiodicity on the drivings, observing that the system eventually thermalizes: It results in featureless random states which can be described by the symmetry of the Hamiltonian. In a system with open edges, we find that both Majorana and fermionic localized edge modes can be engineered with a spatially quasiperiodic potential, in similarity with the case of homogeneous on-site energies. Besides, we demonstrate the possibility of creating multiple Majorana $0$ and $\pi$ modes in a driven setting, even if the underlying static Hamiltonian is in its trivial phase. Lastly, we study the robustness of the Majorana modes against the aperiodicity in the driving period, showing that the ones created via quasiperiodic potential are more robust to the decoherence. Moreover, we find an example where a Majorana mode displays high-robustness, provided that it is chosen from a special point in the topological region.
\end{abstract}

\maketitle

\section{Introduction}

The important role of topology in condensed matter physics was seminally pointed out in the description of the quantum Hall effect~\cite{Thouless82}. The number of conducting edge states is a topological invariant of the system, protected from imperfections that are not sufficient to close the bulk band gaps nor to change the symmetry of the Hamiltonian describing the system, ultimately leading to the striking conductance quantization observed~\cite{Klitzing80}. Recently, the interest in topological states of matter grew enormously~\cite{Hasan10,* Qi11, Shen_book_12,* Asboth_book_16} and one may generically classify the known topological systems in topological insulators and topological superconductors~\cite{Alicea12,* Leijnse12,* Beenakker13,* Aguado17}. In one-dimensional (1D) systems composed by spinless fermions, a representative model of the former is the Su-Schrieffer-Heeger~\cite{SSH_79} (SSH), or equivalent Shockley model~\cite{Shockley39,* Pershoguba12}, whose nearest neighbor (NN) hopping amplitudes are staggered. In the latter, the Kitaev chain, a model manifesting triplet ($p$-wave) superconducting pairing, is another key example~\cite{Kitaev01}. In both cases, they may be interpreted as minimal models describing the topological edge states in experiments involving graphene nanoribbons~\cite{Groning2018} or in both semiconducting nanowires~\cite{Mourik12,* Deng16,* Zhang_MZM_18} or adatom chains on top of a superconductor~\cite{Nadj-Perge14}, respectively.

New phenomena emerge when forcing a quantum system to change with time and one of the simplest ways to do so is to either suddenly or slowly change -- {\em{quench}} -- some of the parameters of the Hamiltonian~\cite{Polkovnikov11, Eisert15}. This is connected to important question of thermalization of an isolated quantum system~\cite{Deutsch91,* Srednicki94,* Kinoshita06,* Rigol07,* Rigol08,* Ziraldo12,* Ziraldo13,* He13,* Vidmar16}. The study of quenches in topological systems showed that the topological order can survive quenches across the topological phase transitions in (infinitely) long systems~\cite{Tsomokos09,* Halasz13,* Rajak14,* Sacramento14,* DAlessio15}. In finite systems however revivals occur, as signaled by the Lochschmidt echo and fidelity~\cite{Gorin06,* Quan06,* Gu10,* Mukherjee12,* Andraschko14,* Sacramento16,* Jafari17,* Dahan17}.

\begin{table}[b!]
\label{Tab:table}
  \caption{Presence of Majorana modes in various settings: T and N refer to Majorana modes created from topological and trivial regions of the undriven Hamiltonian and the star refers to Majorana zero mode created from the flat band point.}  
  \begin{center}
  \ra{1.2}
    \label{tab:table1}
    \begin{tabular}{l c c}
      & & Space: Quasiperiodic \\
      Time & & \\
      \hline
      Static & \phantom{aa}T\phantom{aa} & yes \\ 
      & \phantom{aa}N\phantom{aa} & no \\
      Periodic & \phantom{aa}T\phantom{aa} & yes \\    
      & \phantom{aa}N\phantom{aa} & yes \\ 
      Aperiodic &\phantom{aa}T\phantom{aa} & yes$^*$ \\
       & \phantom{aa}N\phantom{aa} & no \\    
    \end{tabular}
  \end{center}
\end{table}

In parallel, another exciting topic of research is quantum localization, which can be studied either in time-independent systems~\cite{Anderson_58, Aubry_80, Evers_Mirlin_2008, Billy_2008,* Roati_2008} or in time-periodic cases~\cite{Eckardt_2017, Oka18}. The latter, often referred as dynamical localization~\cite{Dunlap_1986}, has been studied in a variety of contexts as, e.g., in two-level systems~\cite{Grossmann_1991}, in quantum kicked-rotors~\cite{Haake, Casati_79,* Fishman_1982,* Moore_1995,* Chabe_2008}, or with a charged particle in a lattice subjected to a sinusoidal force in time~\cite{Dunlap_1986, Lignier_2007,* Eckardt_2009}. It can be generically realized by systems that fail to indefinitely absorb energy from an external drive at some regime. Recently it has been a focus of extensive research also in the context of ergodic properties of driven interacting systems~\cite{DAlessio_2013, DAlessio_2014,* Lazarides_2014,* Regnault_2016,* Lazarides_2014b,* Gritsev_2017,* Nandy17,* Lazarides_2015,* Ponte_2015,* Abanin_2016,* Agarwala_17}.

Here, our goal is to bridge these two aspects, dynamical localization and topological order by studying their interplay. In fact, in the context of quantum driven systems, one can point out the growing interest on the experimental realization of topological states of matter via periodic modulation~\cite{Kitagawa2012,* Rechtsman2013,* Wang2013}. Many other theoretical studies have also highlighted the possibility of creating topological edge states under a periodic drive, in a process dubbed Floquet topological engineering~\cite{Oka09, Kitagawa10, Jiang11, Lindner11, Rudner13, Gomez-Leon13, Thakurathi13, Benito14, Tong13, Kundu2013, Asboth14, Usaj14, Titum15}.

In our case, the starting point is the superconducting Kitaev chain, where the driving in some of the parameters of the Hamiltonian, namely on the chemical potential~\cite{Thakurathi13, Benito14}, on the superconducting phases~\cite{Tong13} or the tunneling~\cite{Benito14}, leads to a multitude of topological (Majorana) modes. Their number specifically depends on the symmetries of the driving and on its frequency. One may also argue that other types of driving can be considered, as the ones that are intrinsically inhomogeneous in real space. An example is the case of quasiperiodic potentials, which can lead to localization either in non-interacting~\cite{Aubry_80, Modugno10,* Kraus12,* Kraus16} or interacting~\cite{Iyer2013,* Mondaini2015} time-independent Hamiltonians. When the quasiperiodic potential varies with time, single-particle~\cite{Qin_14, Cadez17} or many-body~\cite{Bordia_2017} localization are still robust at high-frequencies of the driving.

Thus the questions we address here are: (i) can Majorana modes be engineered with a time-periodic potential that leads to localization? (ii) what are the conditions for their creation? Namely, periods of the driving, range of parameters, etc. (iii) Are other driving protocols, as in the case of aperiodic drivings, robust on the stabilization of Majorana modes? Table~\ref{Tab:table} summarizes the presence or absence of Majorana modes in various settings for quasiperiodic potential (we obtain a similar table with the same entries for homogeneous potentials). We find that one can obtain Floquet edge modes with time-periodic spatially quasiperiodic potentials, but generically these are not robust to aperiodicities in the driving period. The exception is the special case where the Majorana modes are perfectly localized at the edges and the bulk static spectrum is flat. We also demonstrate that Majorana modes created using spatially quasiperiodic driving are more robust to decoherence due to aperiodicity in the driving period than the Majorana modes created using spatially homogeneous driving. This result is in agreement with a recent study, showing that spatial disorder protects topological edge states against the decoherence~\cite{Rieder18,* Sieberer18}. In addition to the investigation of the zero- and finite-quasienergy edge states, we also present detailed analysis of localization properties on the bulk of the spectrum, together with the effects it has on the propagation of initially localized wave-packets, either for homogeneous or quasiperiodic kicks in space.

The paper is organized as follows: We first introduce the model, a kicked Kitaev chain of spinless fermions and describe the basics on Floquet theory in Sec.~\ref{sec:II}. Section~\ref{sec:III} reviews the spatially homogeneous static and time-periodic case. Then we present the detailed analysis of the bulk properties of the spatially quasiperiodic periodically kicked system. Next we study the results of aperiodic kicking in Sec.~\ref{sec:IV}. The time evolution of initialy localized state is presented in Sec.~\ref{sec:V} before focusing on the edge states in Sec.~\ref{sec:VI}. Lastly, we summarize our findings in Sec.~\ref{sec:VII}.

\section{Model and methods} \label{sec:II}

\subsection{Kitaev chain model}

We consider a 1D Kitaev chain model~\cite{Kitaev01} of spinless fermions~\footnote{We note that the static 1D Kitaev chain model is dual to spin $1/2$ $XY$ spin-chain model in a transverse field using Jordan-Wigner transformation~\cite{Lieb61}.}, in a lattice of size $L$ with either open or periodic boundary conditions, whose Hamiltonian reads
\begin{eqnarray}
\hat{H} = \hat{H}_{0J} + \hat{H}_{0 \Delta} + \hat{H}_{0 \mu} + \hat{H}_1,
\label{eq:hamilt}
\end{eqnarray}
where $\hat{H}_{0J} = - \sum_{i} (J_i\,\hat{c}_i^{\dagger} \hat{c}_{i+1}^{\phantom{}} + {\rm H.c.})$ is the kinetic energy, $\hat{H}_{0 \Delta} = - \sum_{i} (\Delta_i\,\hat{c}_{i}^{\dagger} \hat{c}_{i+1}^{\dagger} + {\rm H.c.})$ is the superconducting $p$-wave pairing and $\hat{H}_{0\mu} = - \mu \sum_{i}\hat{c}_i^{\dagger} \hat{c}_{i}$ is the chemical potential and ${\rm H.c.}$ stands for the hermitian conjugate of the preceding terms. The fermionic creation (annihilation) operator at site $i$ is $\hat{c}_i^{\dagger}$ ($\hat{c}_i$); $J_i$ and $\Delta_i$ are the hopping and superconducting $p$-wave pairing between sites $i$ and $i+1$, respectively. Hereafter, we choose homogeneous hoppings ($J_i = J$) and pairings ($\Delta_i=\Delta$), with $J = 1$ setting the energy scale of the problem. The last term in the Hamiltonian $\hat{H}_1 = - \lambda {\sum_{\tau} \delta(t-t_{\tau})} \, \hat{V}$, where $\hat{V} = \sum_{i} \, V_i \, \hat{c}_i^{\dagger} \hat{c}_{i}$, is the potential which is applied onto the system at times $t_{\tau}$ and the integer $\tau$ counts the number of applied kicks. These act as kicks in time by quenching the onsite energies of the lattice whose maximal amplitude is given by $\lambda$. In the periodic case $t_{\tau} = \tau T$ and we employ the Floquet formalism to construct effective time-independent Hamiltonians whose stroboscopic dynamics is equivalent to the one for the original problem.

\subsection{Floquet basics}

The Floquet formalism states that the time-evolution operator describing the dynamics at stroboscopic times of a time-periodic Hamiltonian, $\hat{H}(t+T) = \hat{H}(t)$, is captured by $\hat{U}(nT) = e^{- {\rm i} \hat{H}_{\mathrm{eff}} nT}$, where $ \hat{H}_{\mathrm{eff}}$ is a time-independent Hamiltonian, often referred as the Floquet Hamiltonian.~\cite{Shirley_65,* Sambe_73, Grifoni_98, Bukov_15} Following one period, the time-evolution operator can be written in terms of its eigenstates $|\theta_m\rangle$ and the quasi-energies $\varepsilon_m$, connected to its actual eigenvalues, as $\hat{U}(T) = e^{- {\rm i} \hat{H}_{\mathrm{eff}} T} = \sum_m e^{- {\rm i} \varepsilon_m T} \, | \theta^m \rangle \langle \theta^m |$. \footnote{We note that there is an ambiguity in the definition of the effective Hamiltonian since the Floquet quasi-energies can be shifted by a multiple of $\omega = 2 \pi/T$ without affecting the eigenvalues of $\hat{U}(T)$. The quasi-energy Floquet first Brillouin zone is thus between $-\omega/2$ and $\omega/2$ and $\varepsilon_m T$ is between $- \pi$ and $\pi$.} As will become clear in the following sections, the quasi-energies $\varepsilon_m$ and the corresponding eigenvalues $|\theta_m\rangle$ will provide the basis to study the localization aspects of Eq.\,\eqref{eq:hamilt}.

In general, it is not guaranteed that a closed form of the Floquet Hamiltonian is always obtainable,~\cite{Bukov_15} i.e., if one is able to find an effective time-independent Hamiltonian written in terms of local operators. This is related to the convergence of the Magnus expansion, regularly employed to obtain $\hat{H}_{\rm eff}$ in the high-frequency regime ($T\ll1$). Here, we will deal with a driving protocol that is time-symmetric, i.e., $\hat H(t) = \hat H(T\,-\,t)$, and in the case considered here, the time evolution operator can be written as $\hat{U}(T) = e^{{\rm i} \lambda \hat{V}/2} e^{-{\rm i} \hat{H}_0 T} e^{{\rm i} \lambda \hat{V}/2}$, with $\hat H_0 = \hat{H}_{0J} + \hat{H}_{0\Delta} +\hat{H}_{0\mu}$. Using this simple form, we can write down the Floquet Hamiltonian by making use of the analogue of the Baker-Campbell-Hausdorff (BCH) formula applied to time symmetric problems~\cite{DAlessio_2013}, $\exp{\hat Y} \exp{\hat X} \exp{\hat Y} = \exp\{\hat X + 2 \hat Y - \frac{1}{6}[[\hat X,\hat Y],\hat Y]$ $+ \frac{1}{6}[\hat X,[\hat X,\hat Y]] + \cdots\}$, as
\begin{eqnarray}
 \hat H_{\rm eff} &=& \hat H_0 + \frac{\lambda}{T}\hat V - \frac{T \lambda}{12}[\hat H_0,[\hat H_0,\hat V]] + \nonumber \\
 &+& \frac{\lambda^2}{24}[[\hat H_0,\hat V],\hat V] + \cdots.
 \label{eq:H_eff}
\end{eqnarray}
In the limits of high-frequency $(T\ll1)$ and small kick-amplitudes $(\lambda \ll 1)$, one can truncate the effective Hamiltonian in the first order as (see Appendix~\ref{appendixA})
\begin{equation}
 \hat H_{\rm eff} = \hat H_0 + \frac{\lambda}{T} \hat V.
 \label{eq:floquet_hamil}
\end{equation}
Such expansion tells us that the small period regime is equivalent to the static problem, with an appropriatelly renormalized potential. Although the BCH formula gives us insights on the physics in the small period and kick-amplitude limits, for general parameter values, we use exact diagonalization of the time-evolution operator to probe bulk and edge localization.

\section{Periodically kicked systems}  \label{sec:III}

\subsection{Review of spatially homogeneous kicking}\label{sec:IIIA}

From the high frequency and small kick strength expansion (see Appendix~\ref{appendixA}), we can write down the effective static Hamiltonian, which has the general form
\begin{eqnarray}\label{Heff_hom}
 \hat{H}_{\rm eff} &=& \sum_{i} \, \sum_r \Bigl[ - \bigl( J_r \, \hat{c}_i^{\dagger} \hat{c}_{i+r} + \Delta_r \, \hat{c}_{i}^{\dagger} \hat{c}_{i+r}^{\dagger} + \mathrm{H.c.} \bigr) + \nonumber \\
 & + & \tilde{\mu} \, \hat{c}_i^{\dagger} \hat{c}_{i} \Bigr],
\end{eqnarray}
where the hoppings and pairings between the sites separated by distance $r$ emerge and $\tilde{\mu}$ is the renormalized chemical potential. In the case that the driving kicks are homogeneous in space, $\hat{V} = V\sum_{i} \, \hat{c}_i^{\dagger} \hat{c}_{i}^{\phantom{}}$, it leads to spatially homogeneous parameters $J_r$, $\Delta_r$ and $\tilde\mu$, rescaled by the period, intensities of the kicks and the original parameters in $\hat H_0$.  Generically, the static long-ranged Hamiltonian \eqref{Heff_hom} has been already studied in the literature (see, {\it e.g.}, Refs.~\onlinecite{Niu12, Vodola14, Sacramento15triplet, Alecce17}) and here we briefly revisit some of its main results. 

Considering periodic boundary conditions, after Fourier transforming into momentum space (using $\hat c^\dagger_k = \frac{1}{\sqrt{L}}\sum_j\,e^{{\rm i}k j}\hat c^\dagger_j$), one can rewrite the effective Hamiltonian \eqref{Heff_hom} as
\begin{eqnarray}
 \hat{H}_{\rm eff} = 1/2 \sum_k \, (\hat{c}_k^{\dagger}, \hat{c}_{-k}) {\cal{H}}_k \begin{pmatrix} \hat{c}_k\\ \hat{c}_{-k}^{\dagger} \end{pmatrix},
\end{eqnarray}
where ${\cal{H}}_k = h_x(k) \tau_x + h_y(k) \tau_y + h_z(k) \tau_z$, with $\tau_{\alpha}$ $(\alpha = x,y,z)$ being Pauli matrices in the Nambu space. By solving the Bogoliubov-de Gennes equation ${\cal{H}}_k \psi_k = E_k \psi_k$, where eigenstates $\psi_k = (u_k, v_{-k})^T$, with $u_k, v_{-k}$ being particle and hole coefficients in momentum space, respectively, one obtains the energy-momentum dispersion relation via diagonalization as $E_k^2 = h_x(k)^2 + h_y(k)^2 + h_z(k)^2$, from which the lines in the space of parameters where the gap closes can be trivially determined. In the case of real hopping and pairing coefficients, $h_x(k) = 0$ and the spinless fermion Hamiltonian ${\cal{H}}_k$ belongs to the BDI symmetry class~\cite{Altland97,* Schnyder08,* Kitaev09}, with particle-hole and (generalized) time-reversal symmetries ${\cal{P}} = \tau_x K$ and ${\cal{T}} = K$, respectively and $K$ is the complex conjugation operator. The bulk topological invariant is a winding number~\cite{Read00, Niu12} $W$, which in 1D takes ${\mathbb{Z}}$ values and gives the number of Majorana zero energy edge states in the open system. The winding number can be calculated as $W = 1/(2 \pi) \int {\mathrm{d}k} \, \vartheta(k)$, with $\vartheta(k) = \arctan(h_z(k)/h_y(k))$. For the case of finite range $R$ of hoppings and/or pairings, the highest possible winding number is $R$. An example of an effective static Hamiltonian with next-nearest neighbor (NNN) hoppings and pairings is presented in Appendix~\ref{appendixA}. Recently, there has been an interest also in the infinite range case where the pairing (and/or hopping) decreases either exponentially or as a power-law. In the latter, if the power law exponent is smaller than a threshold value, denoting a regime of extremely long-ranged hoppings (and/or pairings), the so called massive Majorana modes are present~\cite{Vodola14, Viyuela16, Alecce17}.These are described by localized edge states which are gapped from the bulk but their energy is finite in the thermodynamic limit. Finally, in the case of broken time reversal symmetry (complex hopping and/or pairing coefficients) the 1D Hamiltonian ${\cal{H}}_k$ belongs to the symmetry class D~\cite{Altland97,* Schnyder08,* Kitaev09} with a $\mathbb{Z}_2$ valued topological invariant~\cite{Kitaev01}, that can be defined as $\nu = {\rm sign}(h_z(0)\cdot h_z(\pi))$, indicating that the \textit{parity} of the Majorana modes is protected.

Now, in the case of time periodic driving of topological systems, due to the periodicity of the Bloch-Floquet band, additional topological states might appear at the band edge~\cite{Kitagawa10, Lindner11, Jiang11, Rudner13, Gomez-Leon13, Thakurathi13, Tong13, Kundu2013, Benito14, Asboth14, Usaj14, Titum15}, where quasienergy is $\omega/2$ or even within other quasienergy gaps~\cite{Fruchart16, Roy17}. To obtain the correct dynamical bulk-boundary correspondence, the micromotion of the time evolution operator, i.e., its full time evolution throughout the driving cycle, has to be accounted for~\cite{Rudner13, Nathan15, Roy17}. In the case of the  periodically driven Kitaev chain considered here, $0$ or $\pi$ Majorana edge modes can occur~\cite{Thakurathi13, Tong13, Benito14, Molignini18}. In the presence of time reversal symmetry, the corresponding topological invariants are ${\mathbb{Z}} \times {\mathbb{Z}}$ allowing for a multitude of Majorana edge modes~\cite{Thakurathi13, Tong13}. In contrast, for broken time reversal symmetry there can be at most one Majorana of each quasienergy $0$ and $\pi$.

\subsection{Spatially inhomogeneous kicking: localized, critical and extended Floquet states}\label{sec:IIIB}

\begin{figure}[t]
\centering
\includegraphics[width=0.49\textwidth]{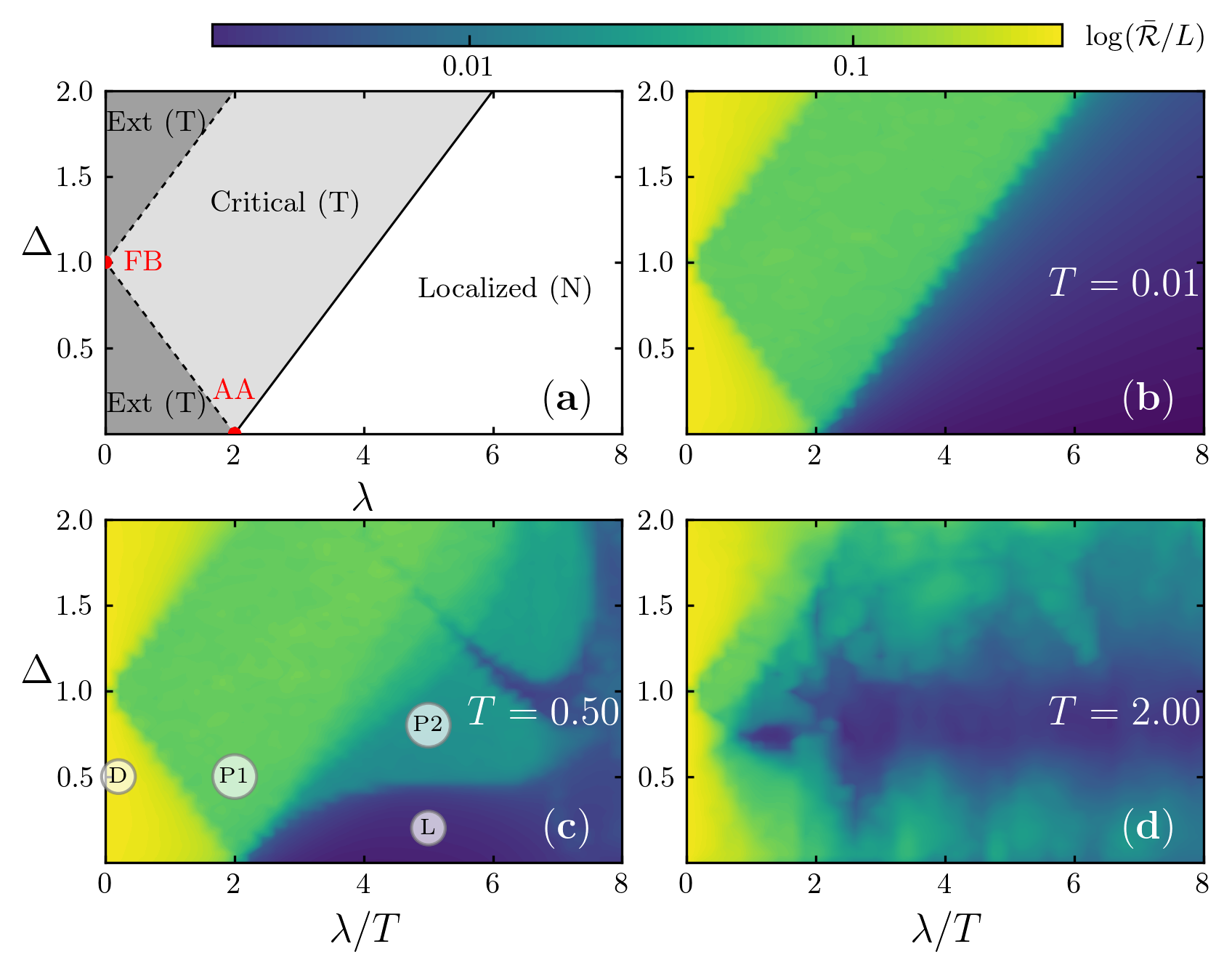}
\caption{Panel (a) shows a schematic phase diagram of spinless fermions in quasiperiodic potential with strength $\lambda$ in the presence of superconducting pairing $\Delta$. There are three distinct phases, where the bulk states are either localized, critical or extended. In the last two cases, localized zero energy edge modes are present in finite (open) systems (marked by T), while the first case is topologicaly trivial (marked by N). Panels (b), (c) and (d) show contour plots of the logarithm of the average mean NPR $\overline{\cal R}/L$ as a function of the strength $\lambda$ of the quasiperiodic kicks on the onsite energies and of the superconducting pairing $\Delta$, at fixed chemical potential $\mu = 0$, for various kicking periods: $T = 0.01, 0.5, 2.0$, corresponding to high, intermediate and low frequencies, respectively. The system size is $L = 500$ and each point is averaged over ten disorder realizations ($r = 10$).}
\label{Fig:IIIB1}
\end{figure}

In this subsection we first review the static system, before presenting our results on the periodically kicked case. Here we consider spatially inhomogeneous potentials, focusing on the quasiperiodic (Aubry-Andr\' e-Harper) potential~\cite{Harper_55, Aubry_80}. This potential is of the form $V_i= \cos(2 \pi \alpha i  + \varphi)$, where we take $\alpha$ as the inverse golden ratio $(\sqrt{5}-1)/2$, which renders its incommensurability with the lattice~\footnote{The choice of inverse golden mean is a standard value used in the studies of quasiperiodic potentials, however we expect to observe the same features with sharp transitions between different phases using other irrational Diophantine numbers, i.e. irrational numbers which are approximated by the rational numbers in the slowest rate. On the other hand if one chooses rational numbers we expect that the boundaries between the phases to smear out considerably, as was demonstrated in the static case in the absence of pairing~\cite{Modugno09}. For non-Diophantine irrational numbers, the smearing of phases is smaller, depending on how well they can be approximated by the rational numbers.}. We have also included an additional phase $\varphi\in [ 0,2\pi)$ that allows  for a ``disorder'' average, thus reducing the statistical and finite-size effects. In the absence of superconducting pairing ($\Delta = 0$), the quasiperiodic potential, for example, arises in the study of free electrons in a 2D square lattice with irrational magnetic fields and it has a striking influence on the spectrum and the eigenstates~\cite{Hofstadter76}. In 1D, contrasting the case of an uncorrelated disordered potential,\cite{Anderson_58} the quasiperiodic case induces a metal-insulator transition at a finite value of the potential strength~\cite{Aubry_80} ($\lambda = 2$). In turn, if this potential is used as a kick~\cite{Leboeuf90,* Artuso94}, a sharp transition occurs at $\lambda/T = 2$ up to intermediate kicking periods ($T \sim 0.5$), where both the critical exponent $\nu$, that describes the behavior of localization near the transition, and the fractal dimension are unaltered in comparison to the static case~\cite{Cadez17}.

\begin{figure}[t]
\centering
\includegraphics[width=0.49\textwidth]{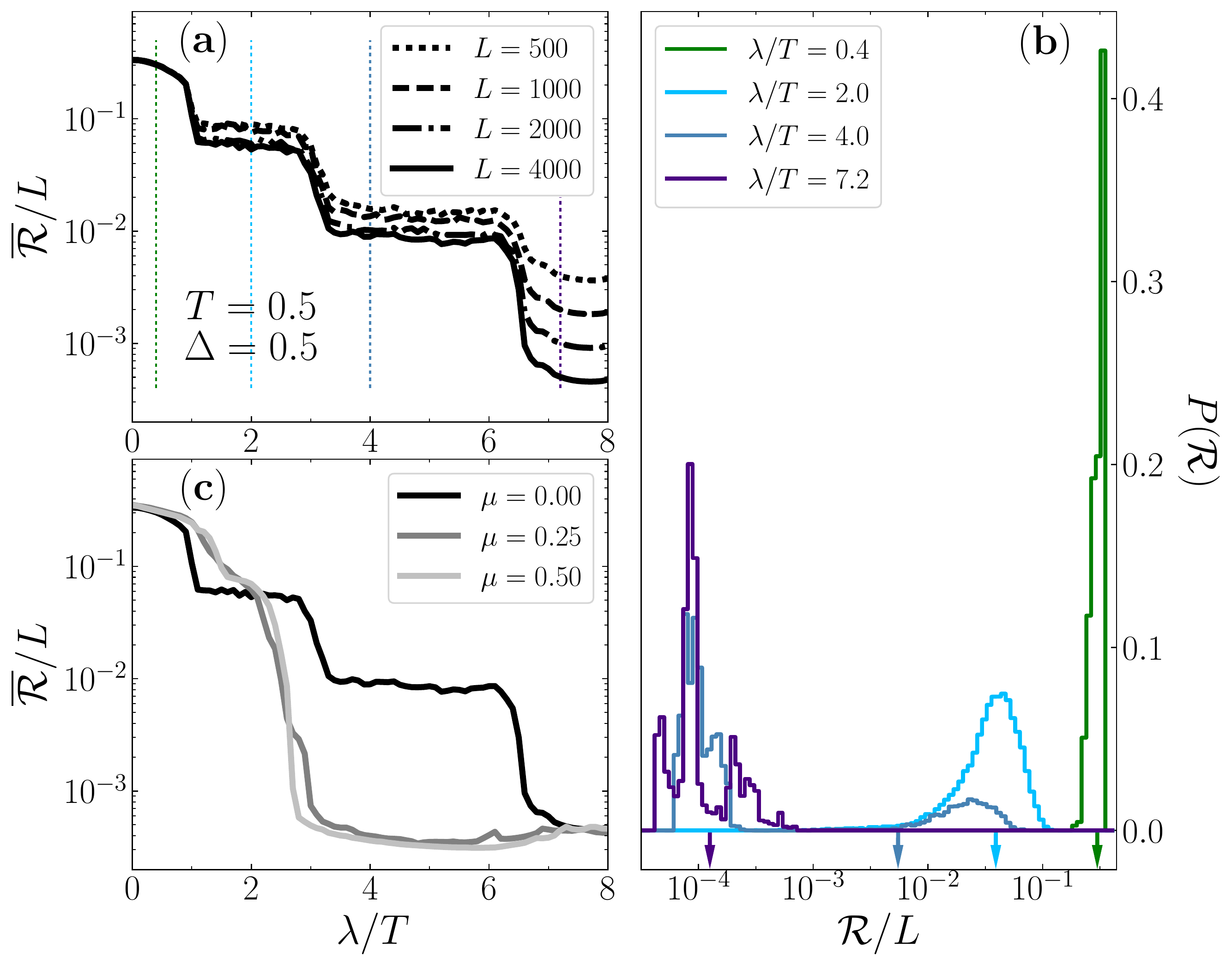}
\caption{Panel (a) shows line cuts of the average mean NPR $\overline{\cal R}/L$ as a function of the strength $\lambda$ of Aubry-Andr\' e-Harper type kicks on the onsite energies for various system sizes at fixed superconducting pairing $\Delta = 0.5$, period of the kicking $T = 0.5$ and chemical potential $\mu = 0$. In (b), the normalized distributions of the PRs, $P({\cal{R}})$, for a few representative kick strengths (marked by the thin dashed lines in panel (a) with corresponding colours). The chosen kicks correspond to phases of delocalized states ($\lambda/T = 0.4$), first plateau of critical states ($\lambda/T = 2.0$), second plateau of a coexistence of critical and localized states ($\lambda/T = 4.0$) and localized states ($\lambda/T = 7.2$). The arrows in panel (b) point to the average mean NPR $\overline{\cal R}$. Panel (c) demonstrates the breakdown of plateaus for nonzero static chemical potential. We used 10 disorder realizations in panels (a) and (c) and a single realization in panel (b). The system size used in panels (b) and (c) are $L = 16000$ and $L = 4000$, respectively.}
\label{Fig:IIIB2}
\end{figure}

Now, turning on the pairing term $\Delta$, we notice that the static Hamiltonian (which, as we previously described, corresponds to the high frequency limit of the kicked case) has been already investigated on what concerns its topological~\cite{DeGottardi13,* DeGottardi13PRB, Cai13} and bulk properties~\cite{Wang16}; the corresponding phase diagram is shown in Fig.~\ref{Fig:IIIB1} (a). There are essentially three phases, classifying the states in the bulk: (i) localized, (ii) critical or multifractal and (iii) delocalized states. If open boundary conditions are used, regions (ii) and (iii) host Majorana edge modes and are topologically nontrivial. We emphasize two special points in the phase diagram: AA highlights the duality point,\cite{Aubry_80} where the metal-insulator transition occurs for $\Delta = 0$, while FB marks the flat band point, where {\em{all}} the states are degenerate (with energy 2) in a periodic, translationally invariant system, whereas two {\em{perfectly}} localized Majorana zero energy states appear in the open system. 

\begin{figure*}[th]
\centering
\includegraphics[width=0.95\textwidth]{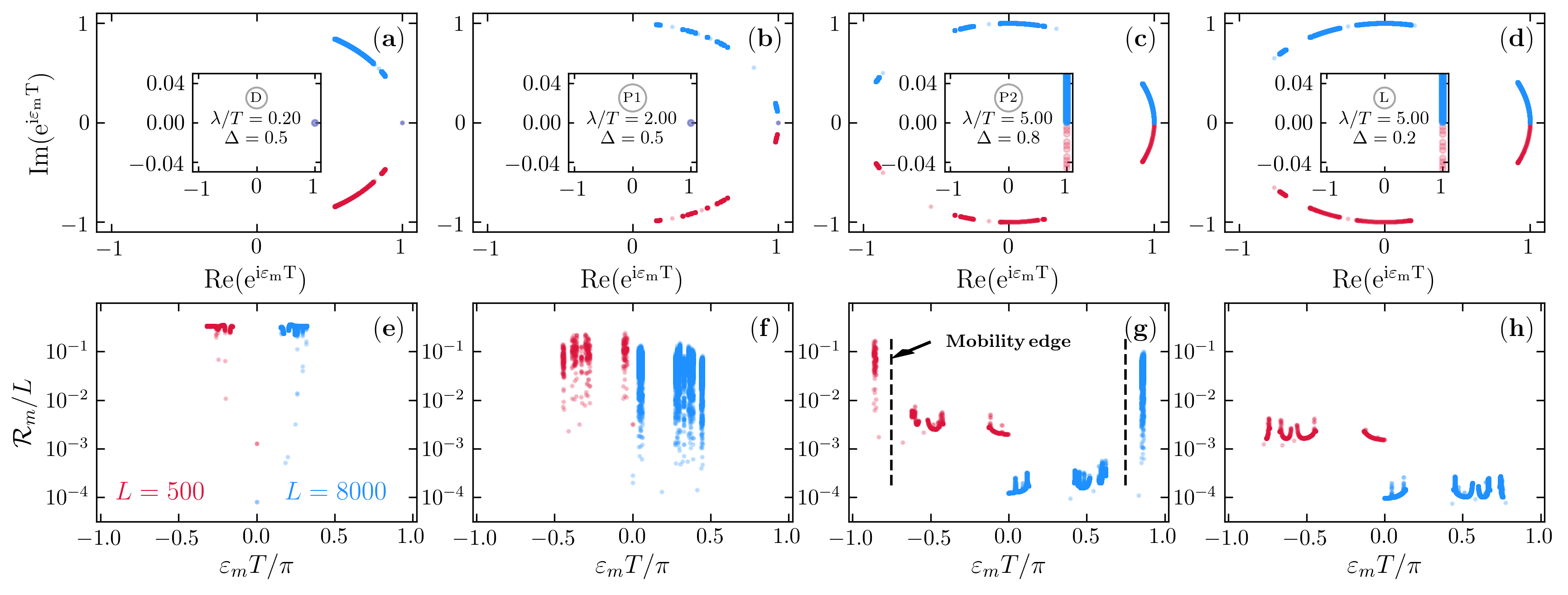}
\caption{(a--d) Real and imaginary part of the eigenvalues of the Floquet operator for representative points of the delocalized (D), critical (first plateau, P1), critical (second plateau, P2) and localized (L) regimes. The insets display a zoom-in of the region of $0$ and $\omega/2$ quasienergies. The corresponding values of the NPR as a function of the states' quasienergies are shown in panels (e-h). The points with $\varepsilon_m T > 0$ ($\varepsilon_m T  < 0$), shown in blue (red) color, correspond to the system size $L = 8000$ ($L = 500$) and we have used $T = 0.5$ and $\varphi = 0$.}
\label{Fig:IIIB3}
\end{figure*}

Localization of the bulk states can be studied via the non-ergodic properties of the system's eigenvalues and eigenvectors.\cite{Haake} To quantify the level of ergodicity, we use the  participation ratio (PR) of the eigenvectors of the Hamiltionian. In that case, we define the PR as ${\cal R}_m = 1/\sum_i (p^m_i)^2$, where the occupation of the Bogoliubov quasiparticle $m$ on site $i$ is given as $p^m_i = |u^m_i|^2 + |v^m_i|^2$ and $u^m_i, v^m_i$ are the corresponding particle and hole coefficients, respectively. The average PR is then $\overline{\cal R} = \overline{\langle {\cal R}_m \rangle_{r}}$, where we first average over all the eigenstates $m$ and then take an average over different disorder realizations $r$. The average PR thus quantifies localization of the eigenvectors in real space. A completely localized state has ${\cal R} = 1$, while a perfectly delocalized state (such as a plane wave) has ${\cal R} = L$. In contrast, critical states scale with the multifractal dimension of the wave-function.\cite{Tang86, Hiramoto_89, Hiramoto_92, Li_Pixley_2016} The three distinct regions of the phase diagram will thus be: (i) localized with the average normalized PR (NPR) $\overline{\cal R}/L \sim {\cal{O}}(1/L)$, (ii) critical with ${\cal{O}}(1/L) < \overline{\cal R}/L < {\cal{O}}(1)$ and (iii) delocalized with $\overline{\cal R}/L \sim {\cal{O}}(1)$.

However, when dealing with a time-periodic problem one instead investigates the level of ergodicity of the eigenstates of the time-evolution operator after one period -- the Floquet operator -- $\hat{U}(T)$, with similar definitions for the PR. We report in Fig.~\ref{Fig:IIIB1}(b) the phase diagram in the regime of high frequency of the kicks ($T=0.01$); one can easily infer its similarity with the case of the static problem~\cite{DeGottardi13,* DeGottardi13PRB, Cai13, Wang16} that persists up to periods $T\sim0.1$. At this period, a second plateau of intermediate mean average NPR, $\overline{\cal R}/L$, starts to emerge at high pairing ($\Delta \sim 2$) in the large $\lambda$ side of the transition line between the critical and localized regions. By increasing the period, the second plateau grows into what was originally a localized region in the high frequency limit. This can be seen in Fig.~\ref{Fig:IIIB1}(c), where we show the phase diagram for period $T = 0.5$. When further increasing the period, the second plateau region moves towards lower values of the pairing $\Delta$ (until it reaches $\Delta = 0$ at $T \sim 0.7$ -- not shown), while simultaneously breaking down at higher $\Delta$, where the average mean NPR indicates localization, as also seen in Fig.~\ref{Fig:IIIB1} (c). This breakdown eventually destroys also the critical region (at $T \sim 1$) and starts moving into the delocalized region as can be seen from Fig.~\ref{Fig:IIIB1} (d) for the period $T = 2$. The absence of a sharp metal-insulator transition for large periods of the kicks was also seen in other contexts, as for instance, when the superconductivity is not present.\cite{Qin_14, Cadez17} Lastly, it is important to point out that finite-size effects do not substantially change this picture: we have observed qualitatively the same phase diagrams for smaller system sizes, down to $L = 20$ (not shown).\footnote{While delocalized states have $\overline{\cal R}/L \sim {\cal{O}}(1)$, the limitation in the distinction of localized and delocalized regions comes from the localization length in the localized regime. However we still clearly see three distinct regions, including the critical region. For the smallest systems considered ($L = 20$) the transitions are though smoothed.} 

To further study the various phases, we focus on the case $T = 0.5$ and consider a line cut in the phase diagram with fixed $\Delta = 0.5$ -- a clear two plateau structure, as shown in Fig.~\ref{Fig:IIIB2}(a) is observed. To understand the nature of the states giving rise to these plateaus, we look into representative points in the phase diagram and instead of checking their corresponding average NPR, we study  the actual normalized distribution $P({\cal{R}})$ in Fig.~\ref{Fig:IIIB2}(b) for a large lattice ($L=16\,000$). Large and small kick amplitudes lead to typically narrow distributions centered around $\overline{\cal R}/L \sim {\cal{O}}(1/L)$ and ${\cal{O}}(1)$, respectively. On the other hand, a kick amplitude which would correspond to the first plateau in Fig.~\ref{Fig:IIIB2}(a), leads to a distribution centered around an average NPR which is not within these previous limits: these are essentially critical  states (across the whole spectrum). Further, the distribution of NPRs associated to the second plateau in Fig.~\ref{Fig:IIIB2} (a) with $\lambda/T=4$ results in a two hump structure, with contributions from critical and localized states.

Next, we focus on four representative points $(\lambda/T, \Delta)$ from the four regions of the phase diagram. These are marked in panel Fig.~\ref{Fig:IIIB1}(c) and correspond to: delocalized states $(0.2, 0.5)$, point D, critical states $(2, 0.5)$, point P1, coexistence of critical and localized states $(5, 0.8)$, point P2, and localized states $(5, 0.2)$, point L. Our goal is to investigate the interplay of localization, as signaled by the NPR of the eigenstates, and the quasienergies $\varepsilon_m$, that can track the presence of topological edge states. For that purpose, we report in Fig.~\ref{Fig:IIIB3}, panels (a-d), the eigenvalues of the Floquet operator $e^{{\rm i}\varepsilon_m T}$ for the four points above defined, considering two different system sizes, $L = 500$ (red) and $L = 8000$ (blue). Given the symmetry on the positive and negative imaginary parts of the eigenvalues, we only display $e^{{\rm i}\varepsilon_m T}$ with positive (negative) imaginary parts for $L=8000\,(500)$.

A general observation is that in all regions there are multiple quasienergy bands, which do not grow with the system size. Points D and P1, consisting of (bulk) extended and critical states, respectively, also host (Majorana) zero quasienergy state and a few localized states, which are due to open boundary conditions and quasiperiodic potential used~\footnote{We have confirmed that no localized states are present in points D and P1 when applying periodic boundary conditions.}. We discuss these further in section~\ref{sec:VI}. The points P2 and L have a band in the region of quasienergy $0$ and due to the larger parameter values, the maximum quasienergy values approach the value of $\omega/2$. When the quasienergies reach $\omega/2$ at even larger values of parameters (or at larger period $T$) the bands start to mix, introducing a new regime of long periods. To confirm the nature of the separate states in the four representative points, we also present in Fig.~\ref{Fig:IIIB3} panels (e-h) the values of the NPRs of each state as a function of the quasienergy. Notice the already mentioned scaling of the localized (point L), critical (points P1 and P2) and delocalized states (points P2 and L). From the values of the NPRs for the point P2 in panel (g), we clearly see a mobility edge between critical and localized states, i.e., quasienergies at the edges of the Floquet band possess NPR which barely decreases with increasing system sizes, depiciting critical states, whereas at the middle of the band they have a noticeable decay, denoting localization of the Floquet eigenstates. This agrees with the analysis of the dual-peaked distribution of PRs given in Fig.~\ref{Fig:IIIB2}(b), representing a point in a similar region of the phase diagram.

\subsection{Critical region and scaling analysis} \label{sec:IIIC}

\begin{figure}[ht]
\centering
\includegraphics[width=0.48\textwidth]{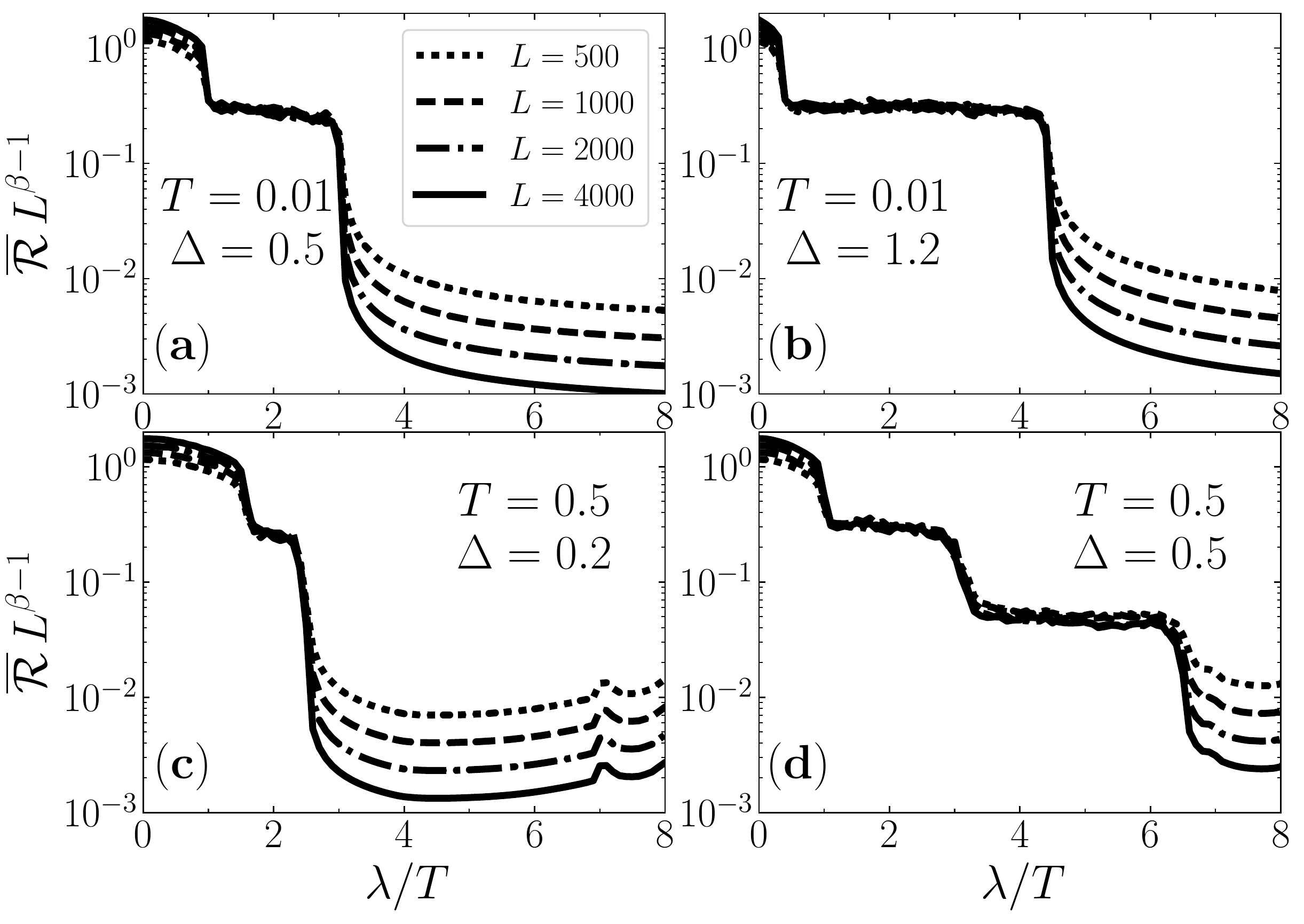}
\caption{Data collapse of the scaled average mean NPR $\overline{\cal R} L^{\beta-1}$ as a function of the strength $\lambda$ of Aubry-Andr\' e-Harper kicks on the onsite energies. We use different system sizes and various pairing magnitudes, considering high ($T=0.01$) and intermediate ($T=0.5$) frequency of the kicks. The chemical potential is set to $\mu = 0$ and the resulting scaling power law exponent is $\beta = 0.2$. We used 10 disorder realizations.}
\label{Fig:IIIC1}
\end{figure}

The critical region, present in both static and kicked cases, is however sensitive to the addition of a finite homogeneous chemical potential $\mu$ in $\hat H_0$. A relatively small potential ($\mu \sim 0.01$) is already sufficient to shrink the plateaus in NPR associated with the critical region, which is finally completely destroyed for larger potentials, as can be seen in Fig.~\ref{Fig:IIIB2} (c) for the cases of $\mu = 0.25, 0.5$. Nevertheless, we will carry out a simple scaling analysis as to argue on the critical behavior of these states, which are neither completely extended nor localized in the case that $\mu = 0$.

We start by recalling that in the static case, a scaling analysis of critical states was recently performed~\cite{Wang16} based on a multifractal analysis.\cite{Tang86, Hiramoto_89, Hiramoto_92} Here, instead, we will focus on the scaling of the average mean NPR, $\overline{\cal R}/L$. We report in Fig.~\ref{Fig:IIIC1} the scaling of this quantity for different values of the period of the kicks ($T=0.01$ and $0.5$) and pairing magnitudes ($\Delta=0.2,\,0.5$ and $1.2$), as a function of the Aubry-Andr\'e-Harper kick strength. For that purpose, we try a scaling form $\overline{\cal R} L^{\beta-1}$, where $\beta$ is a rational number to be adjusted. Indeed, we notice that for the different sets of parameters, the first plateau associated with the presence of critical states across the whole spectrum can be scaled with an exponent $\beta\sim0.2$, at the expense of destroying the collapse for the regions of extended and localized states at small and large $\lambda/T$, respectively. Further, we notice that the second plateau appearing in panel (d), which as we described manifests both critical and localized states, has an almost collapse for this same value of $\beta$. Since in that case the critical states mostly contribute to the average mean NPR [see the arrows in Fig.~\ref{Fig:IIIB2}(b)], we thus expect a close but slightly larger value of $\beta$ in comparison to the purely critical regime, i.e., $\beta \gtrsim0.2$ since in the localized case, $\beta=1$.

\section{Delocalization in aperiodically kicked superconductors} \label{sec:IV}

\begin{figure}[ht]
\centering
\includegraphics[width=0.99\columnwidth]{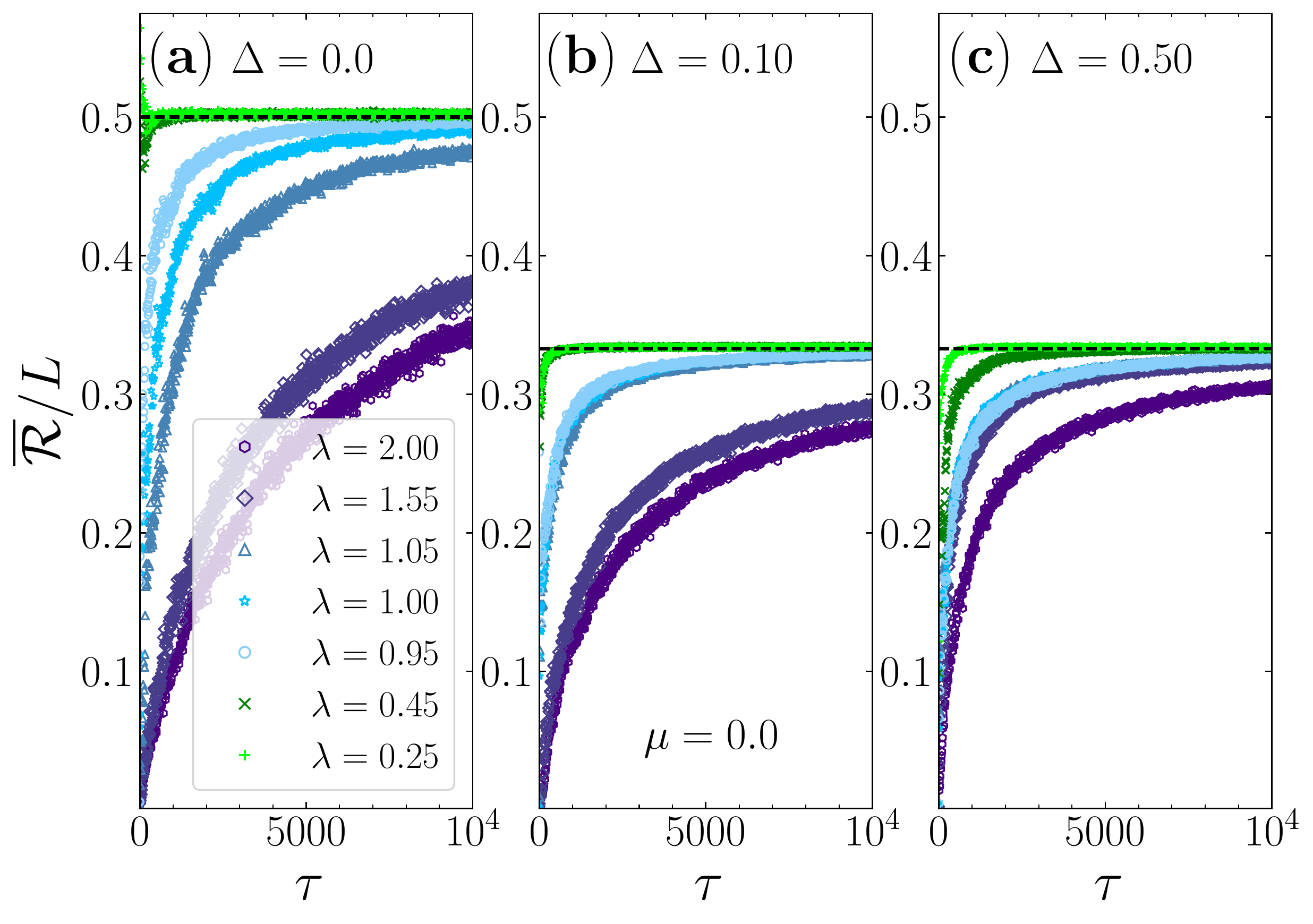} 
\caption{(Color online) Average NPR as a function of the number of kicks in the quantum-kicked Aubry-Andr\'e-Harper model in a lattice with $L = 500$. The (black) dashed horizontal line represents the average NPR of a fully delocalized wave-function obtained from a random matrix belonging to a AI [panel (a)] and BDI [panels (b), (c)] symmetry classes. These refer, respectively, to the cases with zero and finite values of the superconducting pairing $\Delta$. We use the mean periodicity $T = 0.5$ and the maximum aperiodicity $\delta t = T/2$ (see text). A single disorder and time aperiodicity realization was used.}
\label{Fig:IV1}
\end{figure}

In this section we explore the effects of aperiodicity in the driving period on the bulk properies of the Kitaev chain. From an experimental point of view, a small time-aperiodicity is an unavoidable effect and we investigate here the robustness of the different phases we have so far obtained. In fact, the effects of decoherence in non-interacting systems displaying dynamical localization were experimentally studied in the paradigmatic quantum kicked-rotor systems.\cite{Ammann_98,* Klappauf_98, Oskay_2003,* Bitter_2016,* Bitter_2017,* Sarkar17} These experiments demonstrated diffusive behavior of localized wave functions~\cite{Ammann_98,* Klappauf_98} and an unbounded growth of the total energy of the system.\cite{Oskay_2003,* Bitter_2016,* Bitter_2017,* Sarkar17}

\begin{figure*}[th]
\centering
\includegraphics[width=0.95\textwidth]{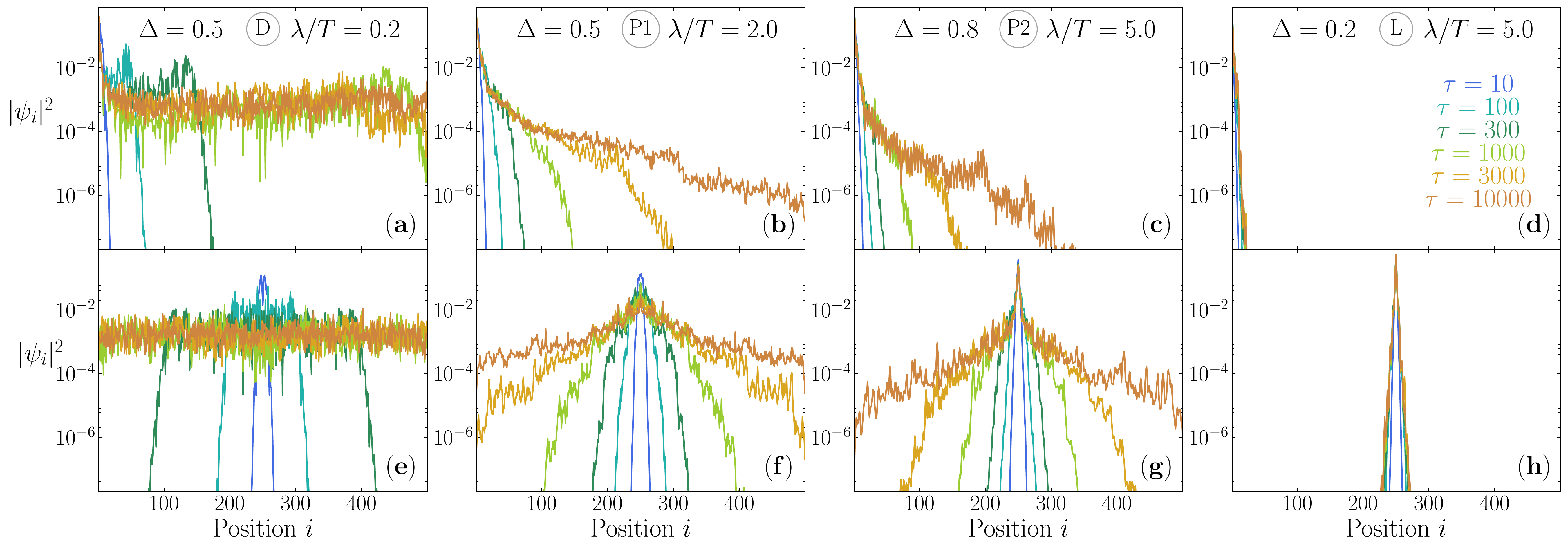}
\caption{The disorder averaged time evolution of initially localized state at the edge (panels a-d) and in the middle of the chain (panels e-h) for the  delocalized, critical (first plateau), critical (second plateau) and localized regime after various numbers of periodic kicks, with period $T = 0.5$. The system size used was $L = 500$ and we averaged over 100 disorder realizations.}
\label{Fig:V1}
\end{figure*}

To study the effects of noise in the time period of the kicks we assume that the time between two consecutive kicks $T_{\tau}$ is a stochastic variable distributed with equal probability between times $T - \delta t$ and $T + \delta t$. The time of ${\tau}$-th kick is then given as $t_{\tau} = t_{{\tau}-1} + T  + \delta t_{\tau}$, where $\delta t_{\tau} \in (-\delta t, \delta t)$ is the aperiodicity for ${\tau}$-th kick and $\delta t < T$ so as to obey causality. The same timing noise scheme was used in the experimental study of a quantum kicked rotor~\cite{Oskay_2003}. The time evolution operator after ${\tau}$ kicks is 
\begin{eqnarray}
 \hat{U}_{\tau} = \hat{U}(T_{\tau}) \hat{U}(T_{{\tau}-1}) \cdots \hat{U}(T_1),
 \label{eq:U_tau}
\end{eqnarray}
with $T_{\tau} = t_{\tau} - t_{{\tau}-1}$ and $\hat{U}(T_{\tau}) = e^{-{\rm i} \hat{H}_0 T_{\tau}} e^{-{\rm i} \lambda \hat{V}}$ and we are interested in the average NPR after a number $\tau$ of aperiodic kicks $\overline{\cal{R}}_{\tau}/L$, obtained from exact diagonalization of Eq.~\eqref{eq:U_tau}.\footnote{Here, unlike in Sec.~\ref{sec:II}, we use non-time symmetric drivings. The protocol used is mostly irrelevant in this context since the difference between one or the other after a driving sequence is merely on a kick with half of the amplitude at the initial and the last drivings.}

A recent numerical study performed by the authors indicates~\cite{Cadez17} that, in the absence of pairing, {\em{any}} nonzero aperiodicity $\delta t$ leads to eventual delocalization at long times and that $\overline{\cal{R}}_{\infty}/L = 0.5$, which is the same as the average NPR of $e^{{\rm i} A}$, where $A$ is a full, random matrix belonging to the Gaussian orthogonal ensemble (GOE) -- or equivalently to the AI symmetry class. From an extensive numerical investigation, we conjectured that this occurs for {\em any} values of period $T$ and kick strength $\lambda$. Here we reach a similar conjecture in the case of non-zero pairing, where $\overline{\cal{R}}_{\infty}/L = 1/3$, which is the same as the average NPR of $e^{{\rm i} A}$, where $A$ is a full, random matrix belonging to BDI symmetry class (chiral GOE). An example demonstrating this behavior is presented in Fig.~\ref{Fig:IV1}, for an average period $T = 0.5$ and multiple values of pairing $\Delta$ and kick strength $\lambda$. Irrespective of whether the starting point belongs to the delocalized, critical or localized regime if the kicks were periodic in time, we observe in all cases that $\overline{\cal{R}}_{\tau}$ goes to the value $1/3$ [marked by dashed lines in panels (b) and (c) of Fig.~\ref{Fig:IV1}] as $\tau \to \infty$. The approach to this asymptotic value is faster for the initially delocalized regime, slower for critical and slowest for the localized one.

\section{Evolution of a localized state} \label{sec:V}

Having probed the aperiodic properties of the kicked problem in the presence of pairing, we return for now to the strictly periodic drivings and focus instead on the transport properties of the Bogoliubov quasiparticle excitations in the Kitaev chain with spatially inhomogeneous kicks. For this, we consider the stroboscopic evolution of an initially localized excitation at the middle and at the edge of an open lattice. On top of being of theoretical relevance, we emphasize  that experiments in optical lattices have used the ability to probe densities with single-site resolution to address coherent single and two particle quantum walks\cite{Preiss15} -- our extra ingredient is the time-periodic and instantaneous quench on the onsite energies.

We start by studying in Fig.~\ref{Fig:V1}, the disorder averaged wave-function evolution for the case of an intermediate period, $T = 0.5$, where we focus on the same points in the phase diagram defined in Fig.~\ref{Fig:IIIB1}(c). The delocalized regime, shown in panels (a) and (e), for initial excitations respectively at the edge and at the middle of the chain,  exhibits ballistic spreading, with the maximum velocity of 0.88 sites per kick and the mean velocity 0.42 site per kick. This can be understood from the following consideration: in the absence of kicks ($\lambda = 0$), the group velocity $v_g(k) = \partial E_k/\partial k$ is $v_g(k) = 2 \bigl[ \mu + 2 \bigl( \Delta^2 - 1 \bigr) \cos(k) \bigr] \sin(k)/E_k$, for the situation of NN hopping and pairing (with periodic boundary conditions). Since an initially completely localized state is a linear combination of all the Bogoliubov excitations of the static Kitaev model, the maximum  group velocity will be connected to the maximum velocity of the spreading, whereas the mean velocity can be estimated as the average group velocity $1/\pi \int_0^{\pi} |v_g(k)| {\mathrm{d}k}$, which gives 1 and 0.64 site per kick, respectively, for the parameters from panels (a) and (e). In fact, we thus observe that the presence of kicks reduces these two values, which may then be recovered in the limit $\lambda \to 0$. Using similar arguments, the revival times after quantum quenches in finite systems were recently studied~\cite{Happola12} in a dual model, the quantum XY model, which provides a connection to the Lieb-Robinson bounds for the light-cone propagation of information in interacting systems.\cite{Lieb72}

Returning to the propagation profile, we note at Fig.~\ref{Fig:V1}(a) that the probability density of the state at the left edge of the system remains at large values after the time evolution with the kicks: this is a clear indication of the formation of a localized edge state in the system for this regime. In direct contrast, if the initial state is initially localized in the middle of the chain, the probability density throughout the lattice eventually becomes completely delocalized, retaining no information about the initial preparation [Fig.~\ref{Fig:V1}(e)].  We observe similar behavior in the (purely) critical regime (point P1), as shown in Figs.~\ref{Fig:V1}(b) and \ref{Fig:V1}(f), although we notice that the spreading slows down considerably; as in the delocalized regime, we observe the presence of the edge state in panel (b). The case of the regime of the second plateau (point P2) signifies the presence of localized states in both the edge and in the middle of the chain [Figs.~\ref{Fig:V1}(c) and \ref{Fig:V1}(g)]: the state after time evolution remains with high probability close to its initial position and only a very slow spreading throughout the chain carries some of its weight. This, however, never happens in the localized regime, where the state always remains exponentially localized, as exemplified in Figs.~\ref{Fig:V1}(d) and \ref{Fig:V1}(h).

\begin{figure}[ht]
\centering
\includegraphics[width=0.99\columnwidth]{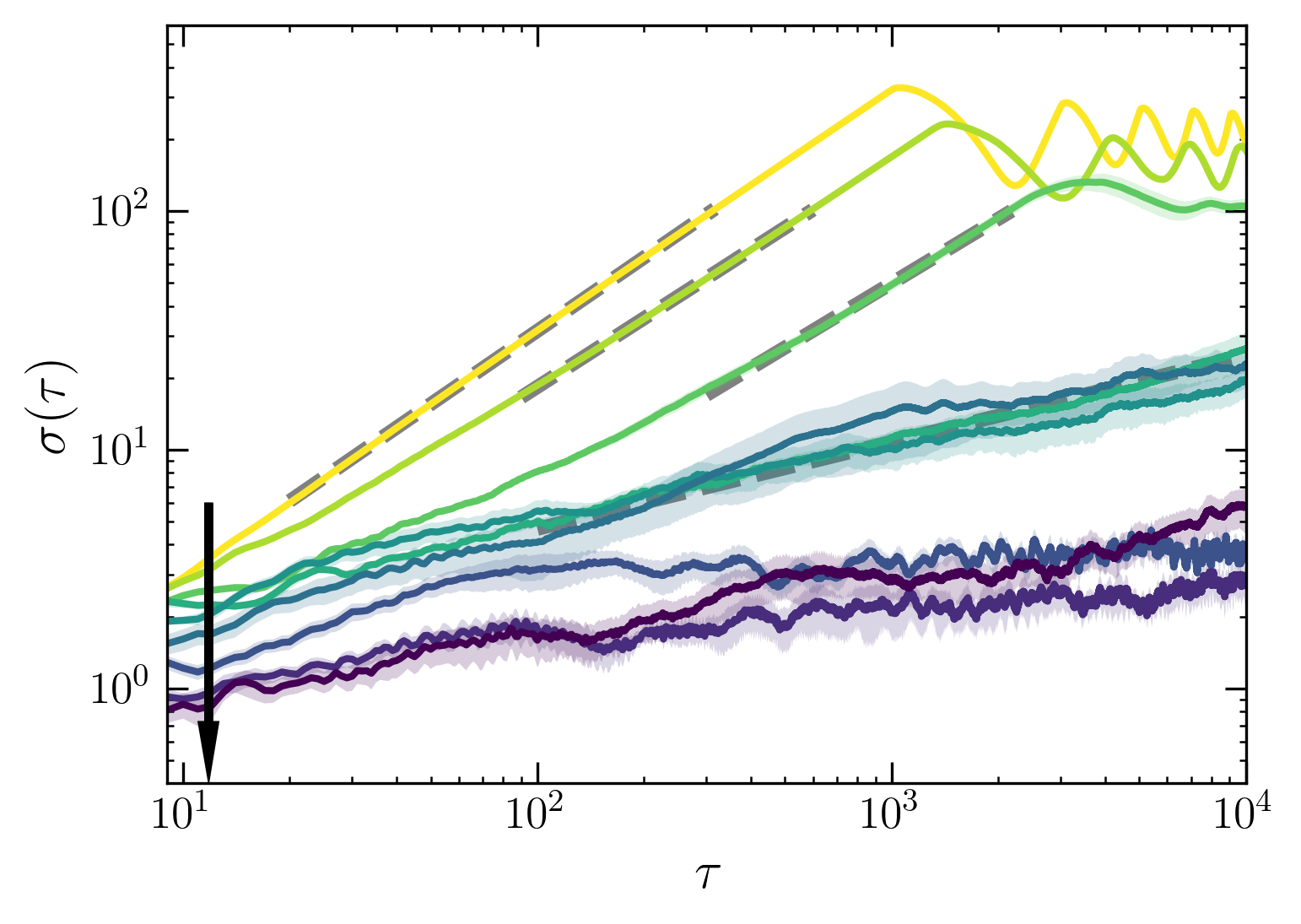}
\caption{Root mean square of the displacement $\sigma$ as a function of number of kicks $\tau$ at fixed $T = 0.5$, $\Delta = 0.5$ and various $\lambda/T = 0, 0.4, 0.8, 1.0, 2.0 , 3.0, 3.6, 4.0, 6.0$ indicated in increasing size by an arrow. Each line is averaged over ten disorder realizations and the shaded area indicates the standard error. Note the log-log scale used, from which we obtain the growth exponents $\gamma$ through the linear fits (indicated by dashed lines for several cases). The system size used was $L = 500$.}
\label{Fig:V2}
\end{figure}

To obtain a more quantitative picture, next we study the wave-function spreading using the root mean square of the displacement, defined as 
\begin{eqnarray}
\sigma(\tau) = \left[\sum_i \left(i - i_0\right)^2 | \psi_i(\tau) |^2\right]^{1/2}, 
\label{eq:sigma}
\end{eqnarray}
where $| \psi_i(\tau) |^2 = |u_i(\tau)|^2 + |v_i(\tau)|^2$ is the density probability at site $i$ after the evolution for $\tau$ kicks of an initially localized state at site $i_0$. The growth of the root mean square displacement is usually of the form $\sigma(\tau) \sim \tau^{\gamma}$, where $\gamma = 1, 1/2$ and $0$ indicates ballistic spreading, diffusion and localization, respectively. The intermediate cases $1/2 < \gamma < 1$ ($0 < \gamma < 1/2$) are denoted as superdiffusion (subdiffusion). 

In Fig.~\ref{Fig:V2}, we quantify the time evolution of $\sigma(\tau)$ in the regime of fixed kick periodicity, $T=0.5$, and pairing, $\Delta = 0.5$, showing the comparison for an increasing magnitude of the kicks. This set of parameters  represent a line cut in the phase diagram Fig.~\ref{Fig:IIIB1}(c), which encompasses different regimes as predicted by the NPRs, also observed in Fig.~\ref{Fig:IIIB3}(a). In the limit $\lambda \to 0$, a free propagation of the state is expected: a ballistic spreading of the initial states' root mean square of the displacement is observed, i.e., $\sigma(\tau) \propto \tau$. By increasing $\lambda$ within the delocalized region, we observe a slight reduction of $\gamma$;  for example, $\gamma \simeq 0.95$ and $0.90$ for the cases $\lambda/T = 0.4$ and $0.8$, respectively. In the critical regime ($1< \lambda/T < 3$) we obtain $\gamma = 0.40 \pm 0.05$ indicating subdiffusive behavior. At even larger $\lambda/T$, corresponding to the region of the second plateau in NPR [Fig.~\ref{Fig:IIIB3}(a)], $\gamma$ decreases even further, until we finally obtain localization and eventually $\gamma \to 0$.

In this analysis of the extraction of the diffusion exponent $\gamma$, we highlight some caveats: First, some care must be taken to avoid the initial transient behavior and the final oscillating regime; the latter is a manifestation of the system's finiteness, where the state has essentially spread over the whole lattice. Second, we note that a single realization (single $\varphi$) behaves differently than the disorder averaged ones considered in Figs.~\ref{Fig:V1} and~\ref{Fig:V2}. To start, there is an intrinsic asymmetry in the position of the expectation value of the time evolved state, originating from the inhomogeneous nature of the quasiperiodic potential. As a consequence, the state propagates more to the side where there is lower potential, which is essentially a single realization aspect. Moreover, scattering centers with high reflectivity are likely to appear, which occur at positions where the ratio of adjacent potential differences is large.

\section{Localized edge states: Majorana \textit{vs} fermionic edge modes}\label{sec:VI}

In this last section, we explicitly explore the formation of the Majorana edge states in time-periodic settings in different parts of the phase diagram, differentiating them from trivial fermionic (Andreev) edge states. Later, we explore their stability under different aperiodic drivings.

\subsection{Periodic kicks}

\begin{figure}[th]
\centering
\includegraphics[width=0.98\columnwidth]{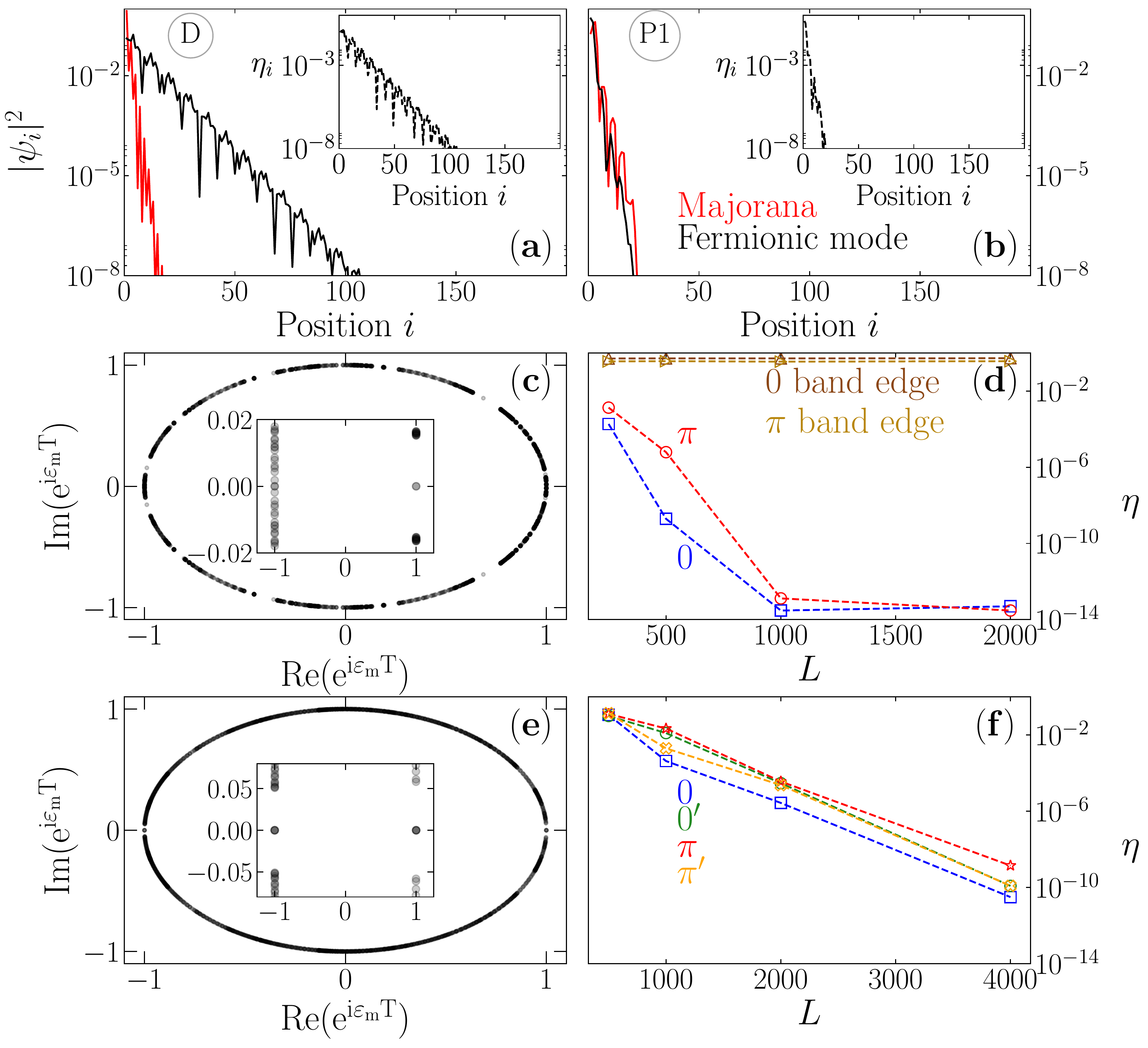}
\caption{Majorana versus normal localized edge modes for the case of delocalized and critical bulk are shown in panels (a) and (b), respectively. Inset shows the difference $\eta_i$ of particle and hole coefficients, as defined in the text. The system size used was $L = 500$ and we zoom in the relevant edge region. Panels (c) and (e) show the quasienergies for two cases of nontopological static Hamiltonian, where multiple quasienergy Majorana modes are created. Self-conjugacy $\eta$ of the Majorana modes from panels (c) and (e) are shown in panels (d) and (f) for one of the Majorana modes in pairs, demonstrating that the corresponding states become self-conjugate exponentially with increasing system size. In contrast we show also two normal states from the band edge in panel (d) which are non self-conjugate. The parameters used in panels (c-d) are $\mu = 0.0, \lambda = 10, T = 2, \Delta = 0.2$ and in panels (e-f) $\mu = 2.5, \lambda = 1, T = 8, \Delta = 0.8$.}
\label{Fig:VI1}
\end{figure}

In the driven case considered here, we examine both the $0$ and $\pi$ quasienergies~\footnote{As is common, we use the terminology and call the Majorana modes at quasienergy $\varepsilon_M = \omega/2$, the $\pi$ Majorana modes, since $\varepsilon_M T = \pi$.} and their corresponding wavefunctions. In Fig.~\ref{Fig:IIIB3}, we showed the full quasi-energy spectrum for representative points of the phase diagram: Note the presence of zero quasienergy states, which are gapped from the extended bulk ones. These are present in the case of delocalized and critical regimes, in panels (a) and (b), respectively. Although suggestive, zero quasienergies are not sufficient to characterize a Majorana mode. Instead, to confirm its nature we also compute the quantity $\eta = \sum_i \eta_i$, with $\eta_i \equiv \bigl| |u_i|^2 - |v_i|^2 \bigr|$ and $u_i$ and $v_i$ being the particle and hole coefficients of the Bogoliubov quasiparticle at site $i$. The value of $\eta$ is vanishing for Majorana states, due to their defining property of being real or self-conjugate ($\gamma_M = \gamma_M^{\dagger}$), from which follows $u_i = v_i^*$.\footnote{Numerically it is not possible to explicitly test this equality, since diagonalization routines introduce arbitrary phases in coefficients $u_i$ and $v_i$. Thus one can only test for equality of moduli.} We thus call quantity $\eta$ self-conjugacy. Examples of the real space probability distributions of the Majorana modes are shown in Figs.~\ref{Fig:VI1}(a) and \ref{Fig:VI1}(b), which can be observed in the regimes where the bulk is delocalized or hosts critical states, respectively. 

Besides Majorana $0$ and $\pi$ quasi-energy states, other localized fermionic edge modes may be present in the system~\cite{Lang12, Ganeshan13, Satija13, Chen17}. In some cases, they can be even more localized than the Majorana states themselves, as exemplified by the lower value of the NPR in Fig.~\ref{Fig:IIIB3}(f) -- their probability distribution is also shown in Figs.~\ref{Fig:VI1}(a) and \ref{Fig:VI1}(b). These are characterized by not possessing 0 or $\pi$ quasienergies, but most importantly by not satisfying the Majorana self-conjugation condition ($\eta\to0$). In practice, one can easily obtain that $\eta$ is essentially zero within machine precision for not so large lattices if dealing with highly localized Majorana states,\footnote{For example, for the case of point D shown in Fig.~\ref{Fig:VI1}(a), we find that already for the system size $L = 50$, the Majorana quasienergy is $\varepsilon_m T \sim10^{-13}$ and corresponding $\eta \sim 10^{-11}$, while both quantities become ${\cal{O}}(10^{-16})$ already for $L = 100$. In contrast for the other localized state quasienergy $\varepsilon_m T = 0.55$ has $\eta = 0.84$, which remains the same with scaling to larger system size.} whereas for the normal fermionic edge modes, they usually possess $\eta_i/|\psi_i|^2 \sim {\cal{O}}(1)$, which ultimately results in a finite $\eta$. We show, in the insets of Figs.~\ref{Fig:VI1}(a) and \ref{Fig:VI1}(b), the site distribution of the self-conjugacy relation: while for the Majoranas they cannot even be represented in the scale for this system size ($L=500$), they are markedly finite for trivial edge modes.

After explicitly characterizing the Floquet Majorana states, we are now in position to tackle an important property of periodically driven systems: starting from a static Hamiltonian whose parameters result in the absence of any topological order, it is possible that the driving induces topologically non-trivial states. This procedure, called Floquet engineering of topological states of matter, has lately received increased attention.\cite{Oka09, Kitagawa10, Lindner11, Jiang11, Rudner13, Gomez-Leon13, Thakurathi13, Kundu2013, Tong13, Benito14, Asboth14, Usaj14, Titum15} For the specific case of the 1D Kitaev chain treated here, this was also recently proposed with different driving protocols.\cite{Thakurathi13, Tong13, Benito14} Here, we show that one can also create multiple topological edge states in the specific case of time-periodic driving with a quasiperiodic potential. Two examples are given in Fig.~\ref{Fig:VI1}, where we present the quasienergies in panels (c) and (e), for a set of parameters such that the underlying static Hamiltonian is trivial. In both cases, gapped quasienergy states close to $0$ and $\pi$ are present, and they exponentially converge to $0$ and $\pi$ with larger $L$'s, a typical characteristic of Majorana states. In the case of panel (c), we notice the manifestation of a pair of states, one at $0$ and one at $\pi$ quasienergies, concomitant with a spatial distribution that is exponentially localized at the edges; they also exponentially approach a perfect self-conjugation with increasing lattice sizes as demonstrated in Fig.~\ref{Fig:VI1}(d). In contrast, typical bulk states right in the vicinity of this quasienergies do not show any change in $\eta$ by increasing system size. In the second example, we show in Fig.~\ref{Fig:VI1}(e) the formation of two pairs of $0$ and $\pi$ quasienergies. In this case, the Majorana end modes are less localized compared to the ones created from the topological regime, but nevertheless exponentially approach $\eta\to0$ with increasing lattice sizes. This is similar to the case of Majorana generation with homogeneous spatial driving,\cite{Thakurathi13} but here with a quench that competes with the localization of the bulk spectrum.

\subsection{Aperiodicity in the driving}

\begin{figure}[ht]
\centering
\includegraphics[width=0.98\columnwidth]{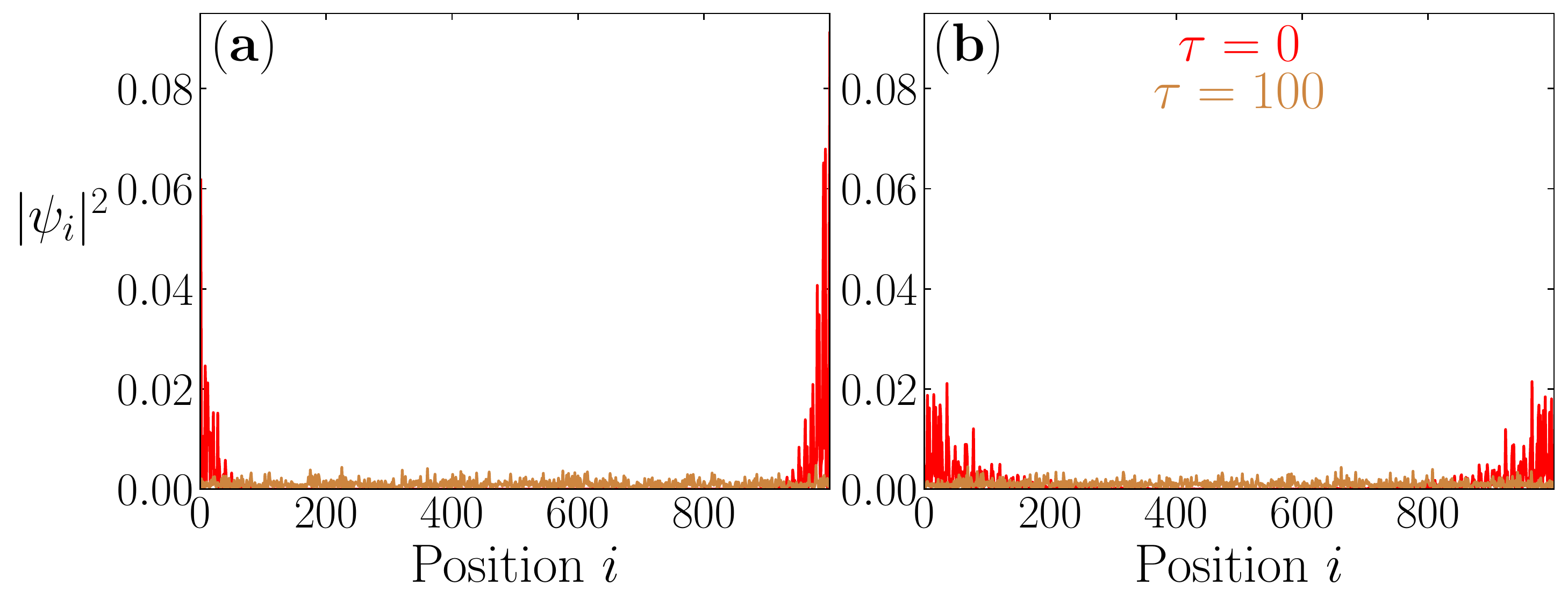}
\caption{Stroboscopic time evolution of a Floquet Majorana mode, engineered from a parent static nontopological phase ($\tau = 0$), under aperiodic driving. Panels (a) and (b) show examples for the case of homogeneous and quasiperiodic kicks in the on-site energies, respectively. The parameters used were $T = 8$, $\mu = 2.5$, $\Delta = 0.8$ and $\lambda = 1.0$ in both panels. The aperiodicity used was $\delta t/T = 0.1$ and the system size $L = 1000$. }
\label{Fig:VI4}
\end{figure}

Finally, an important question concerns the stability of topological states under various types of noises, which were investigated either theoretically~\cite{Balabanov17, Rieder18,* Sieberer18} or experimentally\cite{Jorg17}. For the specific case of Majorana modes in a Kitaev chain this was preliminary studied in Ref.~\onlinecite{Thakurathi13} and expanded in Ref.~\onlinecite{Hu15} using Markovian models of noise.

Here we study the noise due to aperiodicity in the times between successive kicks as introduced in Section~\ref{sec:IV}, i.e., the quenches on the on-site energies all have the same amplitude and phase, but they happen at non-periodic times. For this we track the stroboscopic time evolution of an initial Majorana edge state, Floquet engineered from a parent nontopological static Hamiltonian with a period $T$, and promote time-deviations from this mean period. We start by showing in Fig.~\ref{Fig:VI4} that, generically, an initial Majorana state rapidly decays into the bulk with the application of aperiodic kicks of moderate aperiodicity $\delta t = 0.1\, T$, either in the case where the kicks are spatially homogeneous [Fig.~\ref{Fig:VI4}(a)] or quasiperiodic [Fig.~\ref{Fig:VI4}(b)].

\begin{figure}[t]
\centering
\includegraphics[width=0.98\columnwidth]{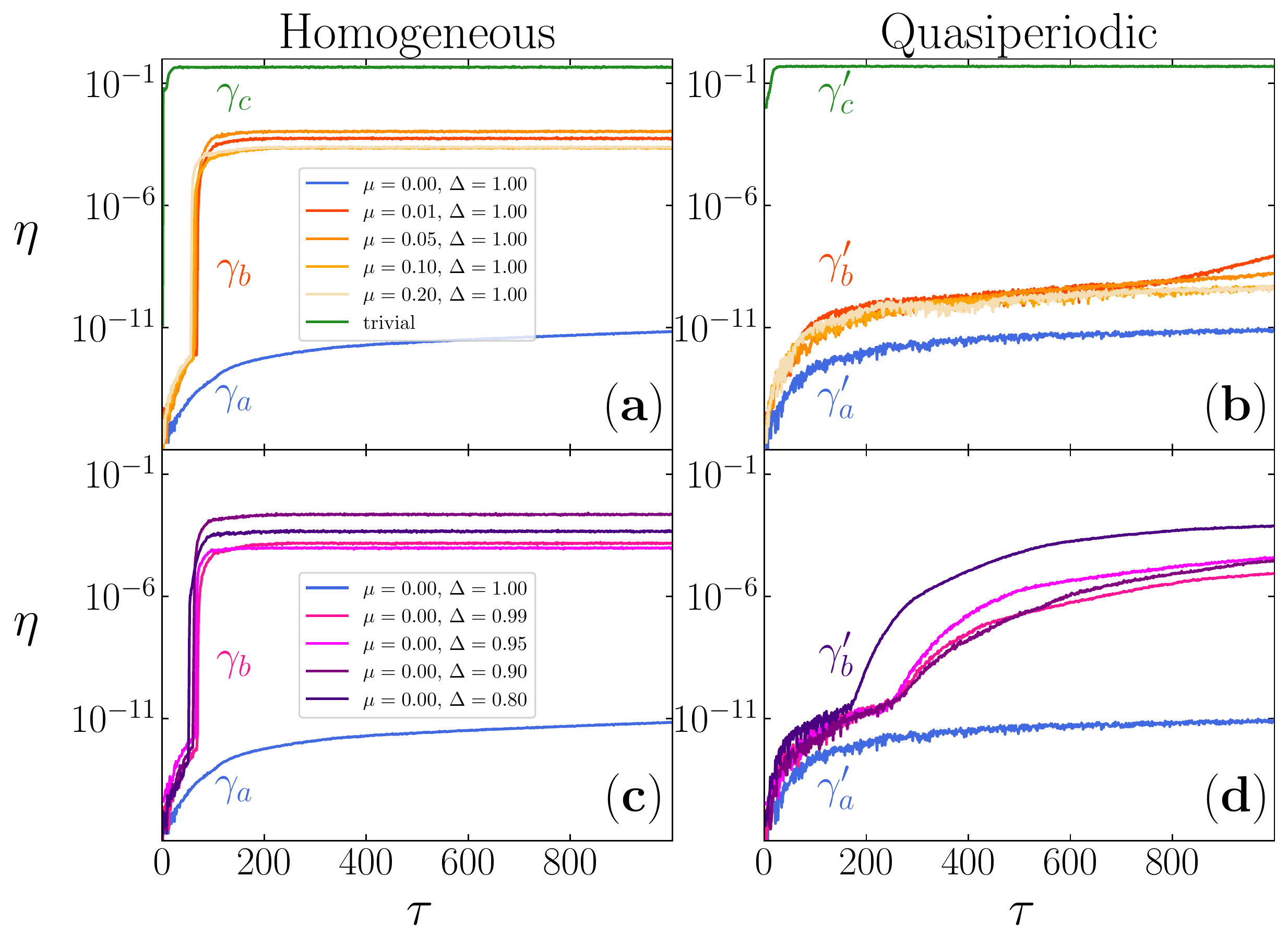}
\caption{Stroboscopic evolution of the self-conjugacy $\eta$ of Majorana modes $\gamma_i$ [$\gamma_i^{\prime}$] as a function of the number of aperiodic kicks in panels (a) and (c) [(b) and (d)] for homogeneous [quasiperiodic] kicks. We present a single Majorana mode $\gamma_a$ [$\gamma_a^{\prime}$] created from the FB point, multiple Majorana modes $\gamma_b$ [$\gamma_b^{\prime}$] from topological regime and a single Majorana mode $\gamma_c$ [$\gamma_c^{\prime}$] from trivial regime (the parameters used here were $T = 8.0, \mu = 2.5, \Delta = 0.8, \lambda = 1.0$). For Majorana modes with $i = a,b$ we used $T = 0.5, \lambda = 0.2$ and in all cases we used $L = 1000, \delta t = 0.1 T$ and a single realization.}
\label{Fig:VI5}
\end{figure}

A more quantitative analysis on the topological nature of the time-evolved state can also be drawn if besides tracking the wavefunction, we study also the evolution of $\eta$, which measures its amount of self-conjugacy. In Fig.~\ref{Fig:VI5}(a)--(c) and \ref{Fig:VI5}(b)-- (d) we show the evolution of $\eta$ for spatially homogeneous and quasiperiodic kicks, respectively. Here we apply aperiodic driving with $\delta t = 0.1 T$ and compare the evolution of three distinct Majorana end modes: (i) Majorana mode $\gamma_a$, which is created from the flat band point ($\mu = 0, \Delta = 1$); (ii) Majorana mode $\gamma_b$, which is created from a generic point in the topological regime; (iii) Majorana mode $\gamma_c$ which is created from a generic point in the trivial regime. Strikingly, the aperiodic time evolution of a Majorana mode $\gamma_a$ (blue) roughly remains self-conjugate up to a large number of kicks~\footnote{We note that the value of $\eta$ grows with the same order of magnitude as does the deviation from normalization of the norm of the time evolved state.}, in contrast to the evolution of the Majorana modes $\gamma_b$ (red, orange, violet) that quickly lose their self-conjugacy, saturating at finite $\eta$ value. Here we observe a difference between the spatially homogeneous and quasiperiodic kicks, namely for the former, the self-conjugacy sharply jumps after about 70 aperiodic kicks to the value of about $10^{-4}-10^{-3}$, where it saturates, while in the latter case self-conjugacy is increasing at a slow, steady rate for fixed $\Delta = 1$ [panel (b)], while it starts increasing faster after about $200$ kicks for the case of fixed $\mu = 0$ [panel (d)]. This demonstrates that Majorana modes induced by quasiperiodic potential are more robust to decoherence against the noise in the aperiodicity of the driving, which is in agreement with a recent study showing that disorder helps to protect topological edge states against the decoherence~\cite{Rieder18,* Sieberer18}. On the other hand, the Majorana mode $\gamma_c$ (green), that starts from the trivial regime, loses self-conjugacy much faster for both drivings. We have also tested larger aperiodicities in the driving, up to the maximum aperiodicity $\delta t = T$ and we observe qualitatively similar results as given in Fig.~\ref{Fig:VI5}.

\begin{figure}[t]
\centering
\includegraphics[width=0.98\columnwidth]{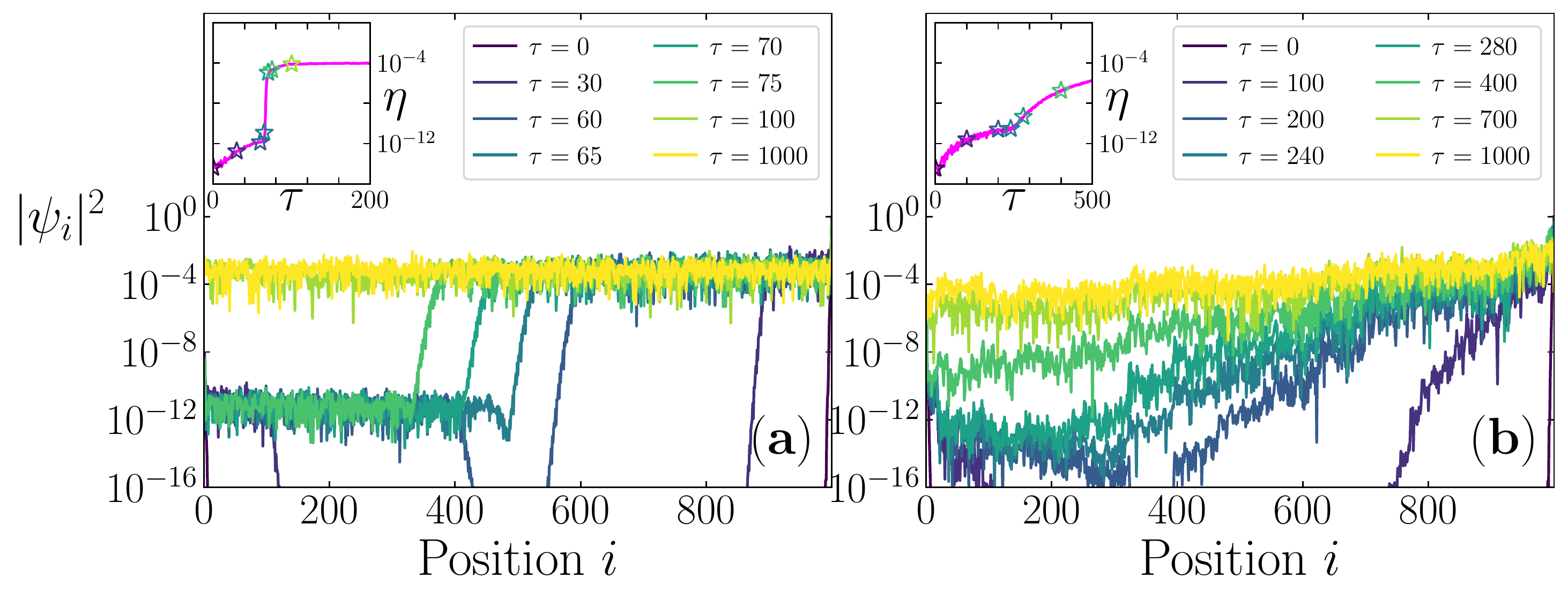}
\caption{Stroboscopic evolution of the Majorana mode $\gamma_b$ after various numbers of aperiodic kicks for the cases of spatially homogeneous and quasiperiodic potential kicks are shown in panels (a) and (b), respectively, demonstating that eventually Majorana modes disperse into the bulk. Insets provide a zoom-in into the evolution of self-conjugacy $\eta$ of Majorana modes as already presented in Fig.~\ref{Fig:VI5}. Star markers depict the number of aperiodic kicks used in the snapshots in the main panels. The parameters used were $L = 1000, \delta t = 0.1 T, T = 0.5, \mu = 0.0, \Delta = 0.95, \lambda = 0.2$.}
\label{Fig:VI6}
\end{figure}

To study the sudden increase of self-conjugacy in more detail we look into snapshots of the  Majorana mode wave-functions at different times, which we show in Fig.~\ref{Fig:VI6} for the case of $\gamma_b$, with $T = 0.5, \mu = 0.0, \Delta = 0.95, \lambda = 0.2$, whose self-conjugacy was already presented in Fig.~\ref{Fig:VI5}. The sudden increase of self-conjugacy occurs when the two edges composing the same Majorana mode hybridize, that is when the wavefunction of the same Majorana mode from both edges begin to overlap. This happens for both homogeneous [Fig.~\ref{Fig:VI5}(a)] or quasiperiodic kicks [Fig.~\ref{Fig:VI5}(b)], albeit in the former at much earlier times due to the larger speed of decaying into the bulk of the edge modes in comparison to the latter.  In fact, the slower increase of self-conjugacy is connected to the nature of the bulk states: in the homogeneous case they are extended, whereas in the quasiperiodic case, for the parameters considered, are critical. Despite eventual complete delocalization of Majorana modes in both cases, we observe that the final self-conjugacy of Majorana modes $\gamma_b$ saturates at a much smaller finite value in comparison to Majorana modes created from a trivial region, $\gamma_c$, indicating that up to a certain level these states retain their distinct property of being self-conjugate.

\section{Summary} \label{sec:VII}

We studied bulk and edge properties of driven 1D Kitaev chain. The driving consisted of instantaneous quenches of the on-site energies, with the main focus on the quasiperiodic modulation of the potential, for both periodic and aperiodic kicks. In the former, we identified three typical driving regimes, the low ($T < 0.1$), intermediate ($T \sim 0.5$) and high period ($T > 1$) ones. In the low period, the time-dependent problem can be mapped onto a time independent effective Hamiltonian whose parameters are renormalized, as obtained from the expansion based on the BCH formula. By deriving the effective Hamiltonian up to the second order in the expansion, we note that the first order term only renormalizes the chemical potential by an additional term which is given by the kick amplitude divided by the period of the driving; in second order, additional terms characterized by NNN pairings and hoppings also appear. In the intermediate regime, the effective Hamiltonian description breaks down. This occurs at about $T \sim 0.1$ provided that other parameters are not much larger than 1 and the kick strength $\lambda$ is of the order of the period $T$ or smaller. Finally, the high period regime occurs when the quasienergy gap at $\omega/2$ closes (quasienergies multiplied by the period become larger than $\pi$) or equivalently $|\varepsilon_m| > \omega/2$ for some state(s) $m$. 

We studied the bulk properties of the spatially inhomogeneous kicked chain via the average mean normalized participation ratio, and by considering the (stroboscopic) time evolution of an initially localized excitation. In the low period regime the phase diagram consists of delocalized, critical and localized regions and the bulk wavefunctions possess different scaling forms with the system size in these regions. A localized excitation spreads nearly ballistically, subdiffusively and does not spread in delocalized, critical and localized region, respectively. In the intermediate regime, an additional phase emerges, hosting a mobility edge in quasienergies between the critical and localized states, which is a new type of mobility edge, contrasting the single particle mobility edges studied previously~\cite{Soukoulis82,* DasSarma88,* Biddle2010,* Ribeiro13,* Ganeshan15,* Roy18} and experimentally measured recently~\cite{Luschen18}. The new phase appears in what was localized region in the low period regime. Lastly, in the high period regime, different quasienergy bands mix and the pure phases break down. The aperiodicity in the driving period leads to the destruction of localization and for long enough drivings the final time evolution operator converges to the random matrix from the corresponding symmetry class, which for the case of $\Delta \ne 0$ ($\Delta = 0$) is BDI (AI).

Finally, we have also studied the edges and demonstrated that in an open system both Majorana and fermionic edge modes are present in delocalized and critical regions of the phase diagram. We have shown that similar to the case of homogeneous kicks, multiple Majorana edge modes can be created by periodic driving with a quasiperiodic potential, thus demonstrating the possibility of Floquet engineering~\cite{Oka09, Kitagawa10, Lindner11, Jiang11, Rudner13, Gomez-Leon13, Thakurathi13, Kundu2013, Tong13, Benito14, Asboth14, Usaj14, Titum15} using spatially quasiperiodic potential. While zero quasienergy Majorana modes can be found in all driving regimes, the $\pi$ Majorana modes occur only in high period regimes, since by definition the $\omega/2$ gap closes at the transition between intermediate and high period regime. By introducing the aperiodicity in the driving, the Majorana modes originating from the trivial regime quickly lose their self-conjugacy. In contrast, the self-conjugacy of Majorana modes from topological regime decays slower and it saturates at a value order of magnitude smaller, despite their complete delocalization, which is an unexpected result. The decay occurs slower for Majorana modes, created by quasiperiodic driving in comparison to spatially homogeneous driving in agreement with a recent study of topological ladder system~\cite{Rieder18,* Sieberer18}. Remarkably, the decay is slowed further in the case of Majorana modes originated from the flat-band point.

\begin{acknowledgments}
T\v C acknowledges fruitful discussions with Chen Cheng, Georg Engelhardt, Panagiotis Kotetes, Guangkun Liu, Tharnier Puel de Oliveira, Marko \v Znidari\v c and the hospitality of Instituto Superior T\'{e}cnico. RM is supported by the National Natural Science Foundation of China (NSFC) Grant No. 11674021 and No. 11650110441 as well as NSAF-U1530401. PDS acknowledges partial support from Funda\c{c}\~ao para a Ci\^encia e Tecnologia (FCT), Portugal  through grant UID/CTM/04540/2013. The computations were performed in the Tianhe2-JK cluster at the Beijing Computational Science Research Center (CSRC).

\end{acknowledgments}

\appendix
\section{High frequency effective Floquet Hamiltonian} \label{appendixA}
Here we derive the effective Floquet Hamiltonian up to (including) the first nested  commutators, as given by the BCH formula in Eq.~\eqref{eq:H_eff}, i.e., the second order in the expansion. By using: (i) standard fermionic anti-commutation relations $\{ \hat{c}_{i}, \hat{c}_{j} \} = 0$ and $\{ \hat{c}_{i}, \hat{c}_{j}^{\dagger} \} = \delta_{i j}$, where $\{ \hat{A}, \hat{B} \} = \hat{A} \hat{B} + \hat{B} \hat{A}$ is an anti-commutator and $\delta_{ij}$ is a Kronecker delta, (ii) elementary commutator relations $[\hat{A} \hat{B}, \hat{C}] = \hat{A} [\hat{B}, \hat{C}] + [\hat{A}, \hat{C}] \hat{B} = \hat{A} \{\hat{B}, \hat{C}\} - \{\hat{A}, \hat{C} \} \hat{B}$, $[\hat{A}, \hat{B} \hat{C}] = [\hat{A}, \hat{B}] \hat{C} + \hat{B} [\hat{A}, \hat{C}] = \{\hat{A}, \hat{B}\} \hat{C} - \hat{B} \{\hat{A}, \hat{C} \}$, $[\hat{A} \hat{B}, \hat{C} \hat{D}] = \hat{A} \{\hat{B}, \hat{C}\} \hat{D} - \hat{A} \hat{C} \{\hat{B}, \hat{D} \} + \{\hat{A}, \hat{C}\} \hat{D} \hat{B} - \hat{C} \{\hat{A}, \hat{D} \} \hat{B}$,  and (iii) explicit forms for the static and kick terms, $\hat{H}_0$ and $\hat{H}_1$, as given in Eq.~\eqref{eq:hamilt}, we notice that we need to calculate the following 4 commutators:
\begin{eqnarray}
{[} \hat{c}_{a}^{\dagger}  \hat{c}_{b}^{\phantom{}} , \hat{c}_{d}^{\dagger} \hat{c}_{e}^{\phantom{}}] & = & \delta_{bd} \, \hat{c}_{a}^{\dagger} \hat{c}_{e}^{\phantom{}} - \delta_{ae} \, \hat{c}_{d}^{\dagger} \hat{c}_{b}^{\phantom{}}, \nonumber \\
{[} {\hat{c}}_{a}^{\dagger}  \hat{c}_{b}^{\dagger} , \hat{c}_{d}^{\dagger} \hat{c}_{e}^{\phantom{}}] & = & - \delta_{be} \, \hat{c}_{a}^{\dagger} \hat{c}_{d}^{\dagger} - \delta_{ae}  \, \hat{c}_{d}^{\dagger} \hat{c}_{b}^{\dagger}, \nonumber \\
{[} \hat{c}_{a}^{\phantom{}}  \hat{c}_{b}^{\phantom{}} , \hat{c}_{d}^{\dagger} \hat{c}_{e}^{\phantom{}}] & = & \delta_{bd}  \, \hat{c}_{a}^{\phantom{}} \hat{c}_{e}^{\phantom{}} + \delta_{ad}  \, \hat{c}_{e}^{\phantom{}} \hat{c}_{b}^{\phantom{}}, \nonumber \\
{[} \hat{c}_{a}^{\dagger}  \hat{c}_{b}^{\dagger} , \hat{c}_{d}^{\phantom{}} \hat{c}_{e}^{\phantom{}}] & = & \delta_{bd} \, \hat{c}_{a}^{\dagger} \hat{c}_{e}^{\phantom{}} - \delta_{be} \, \bigl( \hat{c}_{a}^{\dagger} \hat{c}_{d}^{\phantom{}} - \delta_{ad} \bigr) - \nonumber \\ 
& - & \delta_{ad}  \, \hat{c}_{b}^{\dagger} \hat{c}_{e}^{\phantom{}} + \delta_{ae}  \, \bigl( \hat{c}_{b}^{\dagger} \hat{c}_{d}^{\phantom{}} - \delta_{bd} \bigr).
\end{eqnarray}

The effective second order Hamiltonian is then calculated as
\begin{eqnarray}
 \hat H_{\rm eff} &=& \sum_{i=1}^{L} \biggl[ \Bigl(\tilde{J}_i \, \hat{c}_i^{\dagger} \hat{c}_{i+1} + \tilde{J}^\prime_i \, \hat{c}_i^{\dagger} \hat{c}_{i+2} + \tilde{\Delta}_i \, \hat{c}_{i+1}^{\dagger} \hat{c}_{i}^{\dagger} + \nonumber \\
 &+& \tilde{\Delta}^\prime_i \, \hat{c}_{i+2}^{\dagger} \hat{c}_{i}^{\dagger} + {\mathrm{H.c.}} \Bigr) - \tilde{\mu}_i \, \hat{c}_{i}^{\dagger} \hat{c}_{i} \biggr],
\end{eqnarray}
where $\tilde{J}_i \, (\tilde{J}^\prime_i)$, $\tilde{\Delta}_i \, (\tilde{\Delta}^\prime_i)$ and $\tilde{\mu}_i$ are renormalized nearest (next-nearest) neighbor hopping, nearest (next-nearest) neighbor pairing and onsite potential, respectively. They read
\begin{eqnarray}
 \tilde{J}_i & = & - J_i \Bigl\{ 1 + \nonumber \\
 & + & \lambda \bigl( V_{i+1} - V_{i} \bigr) \bigl[ \lambda \bigl( V_{i+1} - V_{i} \bigr)/2 + T \bigl( \mu_{i+1} - \mu_{i} \bigr) \bigr]/12 \Bigr\}, \nonumber \\
 \tilde{J}^\prime_i & = & T \lambda \bigl[ J_i J_{i+1} \bigl( V_{i+2} - 2 V_{i+1} + V_{i} \bigr) + \nonumber \\
 & - & \Delta_i \Delta_{i+1}^* \bigl( V_{i+2} + 2 V_{i+1} + V_{i} \bigr) \bigr]/12, \nonumber \\
 \tilde{\Delta}_i & = & - \Delta_i \Bigl\{ 1 + \nonumber \\
 & + & \lambda \bigl( V_{i+1} + V_{i} \bigr) \bigl[ \lambda \bigl( V_{i+1} + V_{i} \bigr)/2 + T \bigl( \mu_{i+1} + \mu_{i} \bigr) \bigr]/12 \Bigr\}, \nonumber \\
  \tilde{\Delta}^\prime_i & = & T \lambda \bigl( V_{i+2} + V_i \bigr) \bigl( J_i \Delta_{i+1} + \Delta_i J_{i+1}^*\bigr)/12 \nonumber \\
 \tilde{\mu}_i & = & \mu_i + \lambda/T \, V_i - \lambda T \Bigl[ |J_{i-1}|^2 \bigl( V_{i} - V_{i-1} \bigr) + \nonumber \\
 & - & |J_{i}|^2 \bigl( V_{i+1} - V_{i} \bigr) + |\Delta_{i-1}|^2 \bigl( V_{i} + V_{i-1} \bigr) + \nonumber \\
 & + & |\Delta_{i}|^2 \bigl( V_{i+1} + V_{i} \bigr) \Bigr]/6,
\end{eqnarray}
where we have considered the most general case, where all the terms are in general complex and possess spatial dependence. It is worth mentioning that the first order term in the BCH formula, as identified by Eq.\eqref{eq:floquet_hamil}, merely renormalizes the chemical potential by an addition of a term which is given by the amplitude of the kick divided by the period of the driving. On the other hand, NNN terms in pairing and hoppings arise from the second order term, indicated by the common prefactor $T \lambda$, as well as leads to the renormalization of the nearest neighbor NN terms. To obtain even further range hoppings and pairings higher order terms need to be accounted for. 

\begin{figure}[t]
\centering
\includegraphics[width=0.48\textwidth]{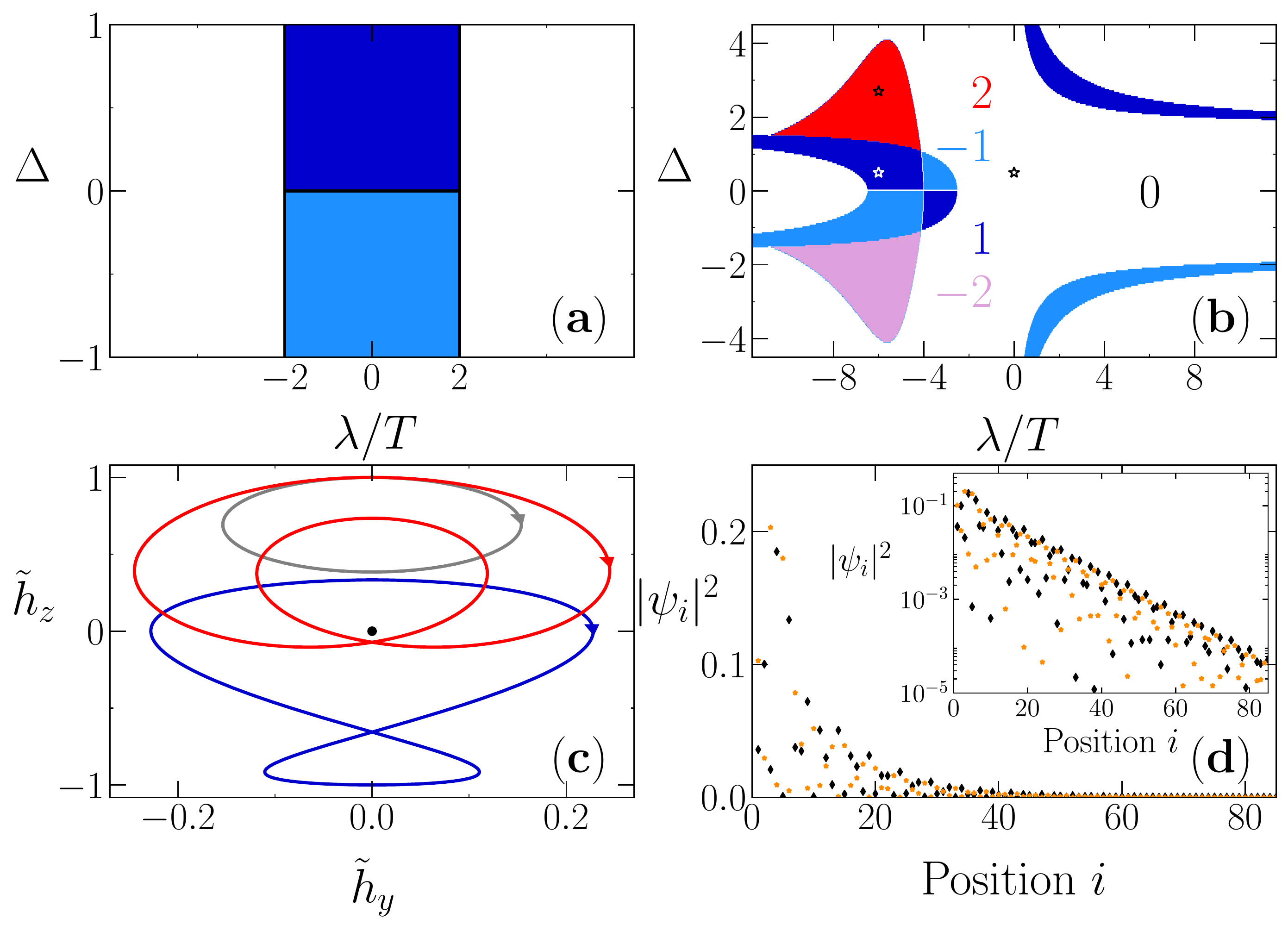}
\caption{Phase diagrams of the effective static Hamiltonian, Eq.~\eqref{Heff_hom_NNN},  shown in the low and intermediate period regimes in panels (a) and (b), respectively -- the $(\mu, T)$ parameters are $(0, 0.01)$ and $(4.5, 0.5)$. In (b), the numbers matching the colors of the regions denote the value of the winding number, while the star symbols depict the examples of different windings, which are shown in panel (c). Panel (c) displays the winding vector, with $\tilde{h}_\alpha = h_\alpha(k)/h_0$  $(\alpha = y,z)$ and normalization $h_0 = {\mathrm{max}}_k (\sqrt{h_y(k)^2 + h_z(k)^2})$. In (d), the two Majorana edge states for $W = 2$ case from panel (c) in a lattice with $L = 1000$; the inset in linear-log scale depicts the exponential localization.}
\label{Fig:AppendixA}
\end{figure}

This general expression may then be simplified when considering a simpler Hamiltonian, with homogeneous NN pairings and hoppings and a constant chemical potential, being driven according to a kick on the site energies that is also spatially homogeneous with amplitude $V$, which we take being $V=1$. In this case, the second order BCH expansion results in the following Floquet Hamiltonian
\begin{eqnarray}\label{Heff_hom_NNN}
 \hat{H}_{\rm eff, h} &=& - \sum_{i=1}^L \, \Bigl[ \bigl( J_1 \, \hat{c}_i^{\dagger} \hat{c}_{i+1} + J_2 \, \hat{c}_i^{\dagger} \hat{c}_{i+2} + \Delta_1 \, \hat{c}_{i+1}^{\dagger} \hat{c}_i^{\dagger} + \nonumber \\
 & + & \Delta_2 \, \hat{c}_{i+2}^{\dagger} \hat{c}_i^{\dagger} + \mathrm{H.c.} \bigr) + \tilde{\mu} \, \hat{c}_i^{\dagger} \hat{c}_{i} \Bigr],
\end{eqnarray}
where the renormalized parameters are $J_1 = J$, $J_2 = - T \lambda | \Delta |^2/3$, $\Delta_1 = \Delta \Bigl[ 1 + \lambda \bigl( \lambda + 2 T \mu \bigr)/6 \Bigr]$, $\Delta_2 = -  T \lambda \Delta {\mathrm{Re}}(J)/3$, $\tilde{\mu} = \mu + \lambda/T - 2 \lambda T | \Delta |^2/3$ . Note that the Hamiltonian \eqref{Heff_hom_NNN} was already studied before (see, e.g., Refs.~\onlinecite{Niu12, Sacramento15triplet}) in the context of static Hamiltonians, but here the model's parameters are expressed in terms of the driven model parameters.

As an exercise of the topological behavior that is manifest in this Floquet Hamiltonian, we start by writing it in momentum space, considering periodic boundary conditions, as
\begin{equation}
 \hat{H}_{\rm eff, h} = 1/2 \sum_k \, (\hat{c}_k^{\dagger}, \hat{c}_{-k}) {\cal{H}}_k \begin{pmatrix} \hat{c}_k\\ \hat{c}_{-k}^{\dagger} \end{pmatrix},
\end{equation}
where ${\cal{H}}_k = - \bigl[\tilde{\mu}/2 + J_1 \cos(k) + J_2 \cos(2 k) \bigr] \tau_z + \bigl[\Delta_1 \sin(k) + \Delta_2 \sin(2 k) \bigr] \tau_y$, with $\tau_{\alpha}$ Pauli matrices in the Nambu space. After diagonalization, it results in the following dispersion $E_k^2 = \bigl[\tilde{\mu}/2 + J_1 \cos(k) + J_2 \cos(2 k) \bigr]^2 + \bigl[\Delta_1 \sin(k) + \Delta_2 \sin(2 k) \bigr]^2$. We report an accurate analysis in Fig.~\ref{Fig:AppendixA}. First, we show that in the high-frequency regime (with $\mu=0$), one recovers the familiar phase diagram of the 1D Kitaev's chain \citep{Kitaev01}, with the gap closings at $\lambda/T = 2$ [see Fig.~\ref{Fig:AppendixA}(a)]. As we have described above, in this regime, the onsite energies get rescaled by the period while the NN hoppings and pairings are unchanged. In contrast, if one studies smaller frequencies [Fig.~\ref{Fig:AppendixA}(b)], the NNN hopping and pairing terms become large enough to give rise to winding numbers $\pm0, \pm1,$ and $\pm 2$. These winding numbers, computed via the transfer matrix method,\cite{DeGottardi13,* DeGottardi13PRB, Alecce17} can be exemplified by the parametric plot of the winding vector $\mathbf{h}(k)=\left(0,h_y(k),h_z(k)\right)$ in Fig.~\ref{Fig:AppendixA}(c), where one can see three cases where this vector rounds around the point $(0,0)$ either 0, 1 or 2 times, denoting winding numbers $W=0,1$ and $2$, respectively. Lastly, as an example of the bulk-boundary correspondence, which relates the winding number with the number of edge states, we report in Fig.~\ref{Fig:AppendixA}(d) the two Majorana edge states for the $W = 2$ case from panel (c), in a lattice with $L=1000$. The inset characterizes the exponential localization of these edge modes.

\bibliographystyle{apsrev4-1}
\bibliography{kicks}

\end{document}